\documentclass[12pt]{article}
\input{epsf}
\usepackage{epsf,cite,epsfig,psfig,float,amssymb,stmaryrd,latexsym}
\usepackage{graphics,psfrag}
\usepackage{a4wide,graphicx}
\usepackage{cite}

\newcommand{\ul}{\underline}
\newcommand{\be}{\begin{equation}}
\newcommand{\ee}{\end{equation}}
\newcommand{\ba}{\begin{eqnarray}}
\newcommand{\ea}{\end{eqnarray}}

\newcommand{\del}{\partial}
\newcommand{\rth}{\frac{1}{\sqrt{3}}}
\newcommand{\rsix}{\frac{1}{\sqrt{6}}}
\newcommand{\sq}{\sqrt}
\newcommand{\fr}{\frac}
\newcommand{\pr}{^\prime}
\newcommand{\ov}{\overline}
\newcommand{\Gm}{\Gamma}
\newcommand{\rw}{\rightarrow}
\newcommand{\rgl}{\rangle}
\newcommand{\De}{\Delta}
\newcommand{\Dp}{\Delta^+}
\newcommand{\Dm}{\Delta^-}
\newcommand{\Dz}{\Delta^0}
\newcommand{\Dpp}{\Delta^{++}}
\newcommand{\Sg}{\Sigma^*}
\newcommand{\Sp}{\Sigma^{*+}}
\newcommand{\Sm}{\Sigma^{*-}}
\newcommand{\Sz}{\Sigma^{*0}}
\newcommand{\X}{\Xi^*}
\newcommand{\Xm}{\Xi^{*-}}
\newcommand{\Xz}{\Xi^{*0}}
\newcommand{\Om}{\Omega}
\newcommand{\Omm}{\Omega^-}
\newcommand{\kp}{K^+}
\newcommand{\kz}{K^0}
\newcommand{\pip}{\pi^+}
\newcommand{\pim}{\pi^-}
\newcommand{\piz}{\pi^0}
\newcommand{\et}{\eta}
\newcommand{\kb}{\ov K}
\newcommand{\km}{K^-}
\newcommand{\kbz}{\ov K^0}
\begin{document}

\title{Baryonic Resonances from Baryon Decuplet-Meson Octet Interaction}

\author{Sourav Sarkar, E. Oset and M.J. Vicente Vacas\\
{\small Departamento de F\'{\i}sica Te\'orica and IFIC,
Centro Mixto Universidad de Valencia-CSIC,} \\
{\small Institutos de
Investigaci\'on de Paterna, Aptd. 22085, 46071 Valencia, Spain}\\
}

\date{\today}

\maketitle
\begin{abstract}
We study $S$-wave interactions of the baryon decuplet with the octet of
pseudoscalar mesons using the lowest order chiral Lagrangian.
In the $S=1$ sector, we find an attractive interaction in the $\Delta K$
channel with $I=1$ while it is repulsive for $I=2$. The attractive interaction
leads to a pole in the second Riemann sheet of the complex plane and is
manifested as a large strength in the scattering amplitude close to the $\Delta
K$ threshold, which is not the case for $I=2$. 

We use the unitarized coupled channel approach to also investigate all 
the other possible values of strangeness and isospin. We find two bound states 
in the $SU(3)$
limit corresponding to the octet and decuplet representations. These are found
to split into eight different trajectories in the complex plane when the $SU(3)$
symmetry is broken gradually. Finally,
we are able to provide a reasonable description for a good number of 4-star
${\frac{3}{2}}^-$ resonances listed by the Particle Data Group. In particular,
the $\Xi(1820)$, the $\Lambda(1520)$ and the $\Sigma(1670)$ states are 
well reproduced. We predict a few other resonances and also evaluate the couplings of the observed resonances to
the various channels from the residues at the poles of the scattering matrix from
where partial decay widths into different channels can be evaluated.
\end{abstract}

\newpage

\section{Introduction} 

The introduction of unitary techniques in a chiral dynamical treatment of the
meson baryon interaction has been very successful. It has lead to good
reproduction of meson baryon data with a minimum amount of free parameters, and
has led to the dynamical generation of many low lying resonances which qualify
as quasibound meson baryon states. These states would qualify as pentaquarks in
the quark picture, although the molecular structure in terms of mesons and baryons
is more appropriate from the practical point of view. At a time when evidence is
piling up for a positive strangeness pentaquark state~\cite{nakano} 
(see~\cite{hyodo} for a thorough list of related theoretical and experimental
papers), it is also worth stating that claims for this multiquark nature of many
non exotic resonances, in the sense that they could be generated dynamically
in a meson-baryon coupled channel approach, have been done
before~\cite{wyld1,dalitz1,frazer,rajasekaran,wyld2}. 

  In this sense although the generation of the $\Lambda(1405)$ in a
multichannel approach had been proved long ago~\cite{dalitz2}, the combined use
of chiral Lagrangians with the Lippmann Schwinger equation in coupled channels
\cite{kaiser,norbert} leads to a successful generation of this resonance and a
fair reproduction of the low energy $K^- N$ data.  The consideration of the
full basis of $SU(3)$ allowed channels in~\cite{angels} made it possible to
reproduce all these data with the lowest order Lagrangian and just one cut off
to regularize the loops.  Further work to make the chiral unitary approach more
systematic was done in~\cite{oller,lutz2,lutz3,lutz4}. In particular,
dimensional regularization was used in~\cite{oller}
which allowed one to make predictions at higher energies.  In this way, 
other resonances, the $\Lambda(1670)$  and the $\Sigma(1620)$, were
generated in the strangeness $S=-1$ sector~\cite{bennhold}, and the finding of another
resonance in the $S=-2$ sector allowed one to associate it to the $\Xi(1620)$
and hence predict theoretically~\cite{bennhold2} the spin and parity of this 
resonance, so far unknown experimentally.  
   In addition the $N^*(1535)$ has also been for long claimed to be another of
these dynamically generated resonances~\cite{siegel,assum,inoue}. Within similar
chiral approaches, the same 
results concerning some of these resonances have  been found in~\cite{nieves}.

  All these works led gradually to a more general result in which a detailed
study of the $SU(3)$ breaking of the problem could show that there are actually
two octets and one singlet of dynamically generated baryons with $J^P=1/2^-$,
coming from the interaction of the octet of pseudoscalar mesons of the pion and
the octet of stable baryons of the proton~\cite{cola,carmen}.  

   The success of these findings motivated further searches and recently it
was found that the interaction of the baryon decuplet of the $\Delta$ and the
octet of mesons of the pion gives rise to a set of dynamically generated
resonances~\cite{lutz}, some of which could be easily identified with existing
resonances and others were more difficult to identify. In addition some peaks of
the speed plot in~\cite{lutz}
appeared on top of thresholds of channels and deserve more thought as to
their physical meaning.  

  In the present work we have taken over the work of~\cite{lutz} and conducted
a systematic search of dynamically generated resonances by looking at poles in
the complex plane.  We have also calculated the residues at the poles, which 
allow us to
determine partial decay widths into different channels and, hence, have more
elements to associate the resonances found to known resonances or new ones so
far unknown.  In addition, we have done a systematic study of the evolution of
the poles as we gradually break $SU(3)$ symmetry from the exactly symmetric case
to the physical case where the difference in physical masses of the mesons and
the baryons lead to a certain amount of $SU(3)$ breaking. 

   The results are interesting.  We confirm the correspondence to known
resonances done in~\cite{lutz} for the most clear cases and discuss others not so
clear, as well as the meaning of the peaks at the thresholds.  In addition
we take into account the width of the $\Delta$ which is too large to be neglected
for a more quantitative study. 
Predictions for new resonances are made, some of which could, in principle, be
easily identified, while others are very  broad and lead to small effects in 
scattering amplitudes,  which could justify why they have not
been reported so far.
  
  In view of our results, particularly concerning partial decay widths of the
resonances found, experimental implications appear which are also discussed in
the paper and should allow further progress in this field.

The paper proceeds as follows: in section~2, we develop the formalism of the
problem. In section~3 a systematic search for the poles in the complex plane is
done and in section~4 a detailed discussion of each resonance is made, concluding
in section~5 with a summary, comments and suggestions.

\section{Formulation}  

    The lowest order term of the chiral Lagrangian relevant for the interaction of the baryon 
decuplet with the octet of pseudoscalar mesons is given 
by~\cite{Jenkins:1991es}\footnote{We use the metric
$g_{\mu\nu}=diag(1,-1,-1,-1)$.} 
\be
{\cal L}=-i\bar T^\mu {\cal D}\!\!\!\!/ T_\mu 
\label{lag1} 
\ee
where $T^\mu_{abc}$ is the spin decuplet field and $D^{\nu}$ the covariant derivative
given by
\be
{\cal D}^\nu T^\mu_{abc}=\del^\nu T^\mu_{abc}+(\Gm^\nu)^d_aT^\mu_{dbc}
+(\Gm^\nu)^d_bT^\mu_{adc}+(\Gm^\nu)^d_cT^\mu_{abd}
\ee
where $\mu$ is the Lorentz index, $a,b,c$ are the $SU(3)$ indices.
The vector current $\Gm^\nu$ is given by
\be
\Gm^\nu=\frac{1}{2}(\xi\del^\nu \xi^\dagger+\xi^\dagger\del^\nu \xi)
\ee
with
\be
\xi^2=U=e^{i\sqrt{2}\Phi/f}
\ee 
where $\Phi$ is the ordinary 3$\times$3 matrix of fields for the pseudoscalar 
mesons~\cite{Gasser:1984gg} and $f=93$ MeV. We shall only study the $S$-wave 
part of the baryon meson interaction and this allows some technical
simplifications. We take for the Rarita-Schwinger fields $T_\mu$ the
representation $Tu_\mu$ from ref.~\cite{bookericson,holstein} with the
Rarita-Schwinger spinor $u_\mu$ given by
\be
u_\mu=\sum_{\lambda, s}{\cal C}(1~\frac{1}{2}~\frac{3}{2}\ ;\ \lambda~s~s_\Delta)\
e_\mu(p,\lambda)\ u(p,s)
\ee
with $e_\mu=(0,\hat e)$ in the particle rest frame, $\hat e$ the spherical
representation of the unit vector $(\lambda=0,\pm 1)$, ${\cal C}$ the Clebsch
Gordan coefficients and $u(p,s)$ the ordinary Dirac spinors
$(s=\pm\frac{1}{2})$. Then eq.~(\ref{lag1}) involves the Dirac matrix elements
\be
\bar u(p\pr,s\pr)\gamma^\nu\ u(p,s)=\delta^{\nu
0}\delta_{ss\pr}+{\cal O}(|\vec p|/M)
\ee
which for the $S$-wave interaction can be very accurately substituted by the
non-relativistic approximation $\delta^{\nu 0}\delta_{ss\pr}$ as done
in~\cite{angels} and related works. The remaining combination of the spinors
$u_\mu u^\mu$ involves
\be
\sum_{\lambda\pr, s\pr}\sum_{\lambda, s}
{\cal C}(1~\frac{1}{2}~\frac{3}{2}\ ;\ \lambda\pr~s\pr
~s_\Delta)\ e^*_\mu(p\pr,\lambda\pr) \
{\cal C}(1~\frac{1}{2}~\frac{3}{2}\ ;\ \lambda~s~s_\Delta)\ 
e^\mu(p,\lambda) \ \delta_{ss\pr}=-1+{\cal O}(|\vec p|^2/M^2)~.
\ee
 Consistently with the non-relativistic approximations done, and the free part of
 the Lagrangian of eq.~(\ref{lag1}) the Rarita Schwinger
propagator undressed from the spinors is the one of an ordinary non-relativistic
particle in quantum mechanics. These
approximations make the formalism analogous to that of~\cite{angels,bennhold2}
regarding the meson baryon loops and the general treatment.

The interaction Lagrangian for decuplet-meson interaction can then be written in 
terms of the matrix 
\be
(\bar T\cdot T)_{ad}=\sum_{b,c}\bar T^{abc} \ T_{dbc}
\ee
as
\be
{\cal L}=3iTr\{\bar T\cdot T\,\,\,\Gm^{0T}\}
\label{lag2}
\ee
where $\Gm^{0T}$ is the transposed matrix of $\Gm^0$, with  $\Gm^{\nu}$ 
given, up to two meson fields, by
\be
\Gm^\nu=\frac{1}{4f^2}(\Phi\del^\nu\Phi-\del^\nu\Phi\Phi).
\ee 

To finalize the formalism let us recall the identification of the $SU(3)$ component
of $T$ to the physical states~\cite{savage,lutz3}:

$T^{111}=\Delta^{++}$, $T^{112}=\rth\Delta^{+}$, $T^{122}=\rth\Delta^{0}$,
$T^{222}=\Delta^{-}$, $T^{113}=\rth\Sigma^{*+}$, $T^{123}=\rsix\Sigma^{*0}$,
$T^{223}=\rth\Sigma^{*-}$,  $T^{133}=\rth\Xi^{*0}$,
$T^{233}=\rth\Xi^{*-}$, $T^{333}=\Omega^{-}$.

Hence, for a meson of incoming (outgoing) momenta $k(k\pr)$ we obtain using
(\ref{lag2}) the 
simple form for the $S$-wave transition amplitudes, similar to~\cite{angels},
\be
V_{ij}=-\frac{1}{4f^2}C_{ij}(k^0+k^{\pr 0}).
\label{poten}
\ee 

The coefficients $C_{ij}$ for reactions with all possible values of
strangeness $(S)$ and charge $(Q)$ are given in Appendix-I. We then
construct isospin $(I)$ states and evaluate the $C_{ij}$ coefficients
using these states, which can be found in Appendix-II.
In the following we present these coefficients for
various possible values of $S$ and $I$ obtained in the present case. For $S=1$
there is only the $\De K$ channel. For $I=1$ and $I=2$ one gets
$C=1$ and $-3$ respectively. According to eq.~(\ref{poten}), positive diagonal 
$C_{ij}$ coefficients indicate attraction in the channel. Hence, there is
attraction in the $I=1$ channel and repulsion in $I=2$.

Next we consider $S=0$. For  $I=\frac{1}{2}$ we have 
the two states $\De\pi$ and $\Sg K$ in coupled channels and in $I=\frac{3}{2}$
we have the $\De\eta$ state in addition to these two. The $C_{ij}$ coefficients
are given in tables~\ref{tabS0I1by2} and \ref{tabS0I3by2} respectively.
In $I=\frac{5}{2}$ the only state is $\De\pi$ for which $C=-3$. 
\renewcommand{\arraystretch}{1.5}

\begin{table}[h]
\begin{center}
\begin{tabular}{c|cc}
\hline
 & $\De\pi$ & $\Sg K$ \\	
\hline 
$\De\pi$ & 5 & 2 \\
$\Sg K$ & & 2 \\
\hline
\end{tabular}
\caption{$C_{ij}$ coefficients for $S=0$, $I=\frac{1}{2}$.}
\label{tabS0I1by2}
\end{center}
\end{table}

\begin{table}[h]
\begin{center}
\begin{tabular}{c|ccc}
\hline
 & $\De\pi$ & $\Sg K$ & $\De\eta$\\	
\hline 
$\De\pi$ & 2 & $\sqrt{\frac{5}{2}}$ & 0 \\
$\Sg K$ & & $-1$ & $\frac{3}{\sqrt{2}}$ \\
$\De\eta$ & & & 0 \\
\hline
\end{tabular}
\caption{$C_{ij}$ coefficients for $S=0$, $I=\frac{3}{2}$.}
\label{tabS0I3by2}
\end{center}
\end{table}

We then consider
$S=-1$. In this case there are the $\Sg\pi$ and $\X K$ states in $I=0$,
the $\De \ov K$, $\Sg\pi$, $\Sg\eta$ and $\X K$ states in $I=1$ and 
the $\De \ov K$ and  $\Sg\pi$ states in $I=2$. The coefficients are given
in tables~\ref{tabS-1I0}, \ref{tabS-1I1} and \ref{tabS-1I2} respectively.

\begin{table}[h]
\begin{center}
\begin{tabular}{c|cc}
\hline
 & $\Sg\pi$ & $\X K$ \\	
\hline 
$\Sg\pi$ & 4 & $\sqrt{6}$ \\
$\X K$ & & 3 \\
\hline
\end{tabular}
\caption{$C_{ij}$ coefficients for $S=-1$, $I=0$.}
\label{tabS-1I0}
\end{center}
\end{table}

\begin{table}[h]
\begin{center}
\begin{tabular}{c|cccc}
\hline
 & $\De\ov K$ & $\Sg \pi$ & $\Sg\eta$ & $\X K$\\	
\hline 
$\De\ov K$ & 4 & 1 & $\sqrt{6}$ & 0 \\
$\Sg\pi$ & & 2 & 0 & 2 \\
$\Sg\eta$ & & & 0 & $\sqrt{6}$ \\
$\X K$ & & & & 1 \\
\hline
\end{tabular}
\caption{$C_{ij}$ coefficients for $S=-1$, $I=1$.}
\label{tabS-1I1}
\end{center}
\end{table}

\begin{table}[h]
\begin{center}
\begin{tabular}{c|cc}
\hline
 & $\De\ov K$ & $\Sg \pi$ \\	
\hline 
$\De\ov K$ & 0 & $\sqrt{3}$ \\
$\Sg \pi$ & & $-2$ \\
\hline
\end{tabular}
\caption{$C_{ij}$ coefficients for $S=-1$, $I=2$.}
\label{tabS-1I2}
\end{center}
\end{table}

For $S=-2$ and $I=\frac{1}{2}$ there are four states that couple with 
each other. These are the $\Sg\ov K$, $\X \pi$, $\X\eta$ and the $\Om K$.
However, only the $\Sg\ov K$ and $\X \pi$ states couple in $I=\frac{3}{2}$.
The corresponding $C_{ij}$ coefficients are given in tables~\ref{tabS-2I1by2}
and \ref{tabS-2I3by2} respectively. 
\begin{table}[hbt]
\begin{center}
\begin{tabular}{c|cccc}
\hline
 & $\Sg\ov K$ & $\X \pi$ & $\X\eta$ & $\Om K$\\	
\hline 
$\Sg\ov K$ & 2 & 1 & 3 & 0 \\
$\X\pi$ & & 2 & 0 & $\frac{3}{\sqrt{2}}$ \\
$\X\eta$ & & & 0 & $\frac{3}{\sqrt{2}}$ \\
$\Om K$ & & & & 3 \\
\hline
\end{tabular}
\caption{$C_{ij}$ coefficients for $S=-2$, $I=\frac{1}{2}$.}
\label{tabS-2I1by2}
\end{center}
\end{table}

\begin{table}[hbt]
\begin{center}
\begin{tabular}{c|cc}
\hline
 & $\Sg\ov K$ & $\X \pi$ \\	
\hline 
$\Sg\ov K$ & $-1$ & 2 \\
$\X \pi$ & & $-1$ \\
\hline
\end{tabular}
\caption{$C_{ij}$ coefficients for $S=-2$, $I=\frac{3}{2}$.}
\label{tabS-2I3by2}
\end{center}
\end{table}

We then consider the case $S=-3$. In this case the $\X\ov K$ couples with
the $\Om \eta$ state in $I=0$ and with $\Om \pi$ in $I=1$. The coefficients are
given in the tables~\ref{tabS-3I0} and \ref{tabS-3I1} below.
\begin{table}[hbt]
\begin{center}
\begin{tabular}{c|cc}
\hline
 & $\X\ov K$ & $\Om \eta$ \\	
\hline 
$\X\ov K$ & 0 & 3 \\
$\Om \eta$ & & 0 \\
\hline
\end{tabular}
\caption{$C_{ij}$ coefficients for $S=-3$, $I=0$.}
\label{tabS-3I0}
\end{center}
\end{table}

\begin{table}[h]
\begin{center}
\begin{tabular}{c|cc}
\hline
 & $\X\ov K$ & $\Om \pi$ \\	
\hline 
$\X\ov K$ & $-2$ & $\sqrt{3}$ \\
$\Om \pi$ & & $-2$ \\
\hline
\end{tabular}
\caption{$C_{ij}$ coefficients for $S=-3$, $I=1$.}
\label{tabS-3I1}
\end{center}
\end{table}

Finally, for $S=-4$, there is only the state $\Om \ov K$ with $I=\frac{1}{2}$
for which $C=-3$.

The $SU(3)$ decomposition of the ${\frac{3}{2}}^-$ states obtained from the 
combination of a decuplet and a octet is given by
\be
10\otimes 8=8\oplus 10\oplus 27\oplus 35.
\label{decomp}
\ee
In order to perform a systematic study of possible dynamically generated resonances, it is
instructive to evaluate the strength of the potential in each of the 
above representations. To this end the matrix elements of 
the transition potential or equivalently the $C_{ij}$ coefficients calculated above
are now projected on to the 
basis of $SU(3)$ states, 
\be
C_{\alpha\beta}=\sum_{i,j}\langle i,\alpha\rangle C_{ij}\langle j,\beta\rangle  
\ee
where $\alpha,\beta$ denote the basis of $SU(3)$ states on the right hand side of
eq.~(\ref{decomp})
and $i,j$ represent the states in the isospin basis.  
Using the $SU(3)$ Clebsch-Gordan coefficients $\langle i,\alpha\rangle$,
we obtain 
\be
C_{\alpha\beta}=diag(6,3,1,-3)
\ee
in the order of 8, 10, 27 and 35. This indicates that there is strong attraction 
in the octet channel followed by the decuplet. There is also weak attraction in
the 27-plet and a repulsion in the 35-plet. This was already noted
in~\cite{lutz}.

Having thus obtained the matrix $V$ of  eq.~(\ref{poten}), it is
used as the kernel of the Bethe Salpeter equation  to
obtain the transition matrix fulfilling exact unitarity 
in coupled channels~\cite{angels}.
This leads us to the matrix equation
\be
T=(1-VG)^{-1}V.
\label{LS}
\ee
The same result is also obtained using the N/D approach of unitarization~\cite{oller}.
In eq.~(\ref{LS}), $V$ factorizes on shell~\cite{angels,oller} and the diagonal matrix $G$ stands for the loop function
of a meson and a baryon. For a cut off regularization the expression for
$G_l$ is given by~\cite{angels}
\begin{eqnarray}
G_{l} &=& i \, \int \frac{d^4 q}{(2 \pi)^4} \, \frac{M_l}{E_l
(\vec{q})} \,
\frac{1}{k^0 + p^0 - q^0 - E_l (\vec{q}) + i \epsilon} \,
\frac{1}{q^2 - m^2_l + i \epsilon} \nonumber \\
&\rightarrow& \int_{|\vec q|<q_{max}} \, \frac{d^3 q}{(2 \pi)^3} \, \frac{1}{2 \omega_l
(q)}
\,
\frac{M_l}{E_l (\vec{q})} \,
\frac{1}{p^0 + k^0 - \omega_l (\vec{q}) - E_l (\vec{q}) + i \epsilon}
\label{propcutoff}
\end{eqnarray}
with $k^0$, $p^0$ the initial meson and baryon energies, and $\omega$, $E$ 
the meson and baryon intermediate energies.   
 
In the dimensional regularization scheme one has~\cite{oller},
\begin{eqnarray}
G_{l}&=& i \, 2 M_l \int \frac{d^4 q}{(2 \pi)^4} \,
\frac{1}{(P-q)^2 - M_l^2 + i \epsilon} \, \frac{1}{q^2 - m^2_l + i
\epsilon}  \nonumber \\ 
&=& \frac{2 M_l}{16 \pi^2} \left\{ a_l(\mu) + \ln
\frac{M_l^2}{\mu^2} + \frac{m_l^2-M_l^2 + s}{2s} \ln \frac{m_l^2}{M_l^2} 
\right. \nonumber \\ & &  \phantom{\frac{2 M}{16 \pi^2}} +
\frac{q_l}{\sqrt{s}}
\left[
\ln(s-(M_l^2-m_l^2)+2 q_l\sqrt{s})+
\ln(s+(M_l^2-m_l^2)+2 q_l\sqrt{s}) \right. \nonumber  \\
& & \left. \phantom{\frac{2 M}{16 \pi^2} +
\frac{q_l}{\sqrt{s}}}
\left. \hspace*{-0.3cm}- \ln(-s+(M_l^2-m_l^2)+2 q_l\sqrt{s})-
\ln(-s-(M_l^2-m_l^2)+2 q_l\sqrt{s}) \right]
\right\},
\label{propdr}
\end{eqnarray}
where $\mu$ is the scale of dimensional regularization. Changes in the 
scale are reabsorbed in the subtraction constant $a(\mu)$ so that the results 
remain scale independent. In eq.~(\ref{propdr}), $q_l$ denotes the three-momentum of the meson or baryon in
the centre of mass frame and is given by
\be
q_l=\frac{\lambda^{1/2}(s,m_l^2,M_l^2)}{2\sqrt{s}}
\ee
where $\lambda$ is the triangular function and $M_l$ and $m_l$ are the masses of
the baryons and mesons respectively. 

As discussed in Ref.~\cite{oller}, a straightforward comparison between the
propagators with the two regularization schemes as given by
eq.~(\ref{propcutoff}) and (\ref{propdr}) yields a relation between the
cut-off momentum $q_{max}$ and the subtraction constant $a$. Considering a
natural size value for $q_{max}$ to be in the region of the $\rho$ mass, this
relation gives a value of $a\sim -2$. Following~\cite{oller}, we use the 
values $q_{max}=\mu=700$ MeV and $a=-2$ in this study. With this choice, the
values of $G$ in the two approaches are very similar around threshold and in 
a relatively large range of energies. However, because of its good
analytical properties, we have used $G$ given by eq.~(\ref{propdr}) for
searching poles in the different Riemann sheets. The cut-off propagator
(\ref{propcutoff}) is used
only in cases where the width of the baryon is considered.

A strict chiral counting would require that eq.~(\ref{propdr}) behaved as $O(p)$,
but the full relativistic form contains the term $a_l(\mu) + \ln
\frac{M_l^2}{\mu^2}$ which is not homogeneous in the external four-momenta of
the mesons. This problem appearing when calculating loops relativistically is
well known~\cite{Gasser:1987rb}. It is possible to 
obtain expressions which
behave as $O(p)$ and one of them is given in Ref.~\cite{lutz}.
For what respects our work, we refer to Ref.~\cite{oller} where the problem is
discussed and it is concluded that a matching to $\chi PT$ at low energies can
be equally done in spite of this. 

Another difference of the present approach
with respect to the one in Ref.~\cite{lutz} concerns the choice in~\cite{lutz}
of an energy where the loop function $G(s)$ vanishes and hence $T\equiv V$, as
a consequence of which the $T$-matrix manifests the crossing symmetry of the
kernel $V$ at this energy. This energy is chosen in~\cite{lutz} at the mass of
the baryons, an educated guess which makes the scheme more symmetric with
respect to extrapolations above and below that energy using the unitarization
schemes~\cite{lutz_talk}, and which leads to acceptable results when compared to
experiment. In our scheme we choose the subtraction constant in eq.~(\ref{propdr})
as the one found in~\cite{bennhold} which gives good results in the study of the
$K^-p$ interactions, and we find that it also leads to sensible results in the
present situation. In fact, our scheme and others in the
literature~\cite{angels,oller} also have the $G$ function vanishing at an energy
similar to the one of Ref.~\cite{lutz}. For instance, in the present case
for the $\De\pi$ loop we find that $G(s)$ vanishes for $\sqrt{s}\simeq 750$ MeV.
 
Another difference between Ref.~\cite{lutz} and the present work is the
different approaches followed in order to project over the $s$-wave, which is
the one we have considered in the present case. In Ref.~\cite{lutz}, the
authors use projection operators which project over the $s$, $p$ and $d$ waves.
The exact projection over $s$-wave within the scheme followed here is done in
Ref.~\cite{bennhold}. The expression used in eq.~(\ref{poten}) for the kernel is a non-relativistic
approximation done in Ref.~\cite{angels} which neglects terms of $O(q/2M)^2$.
The expression for $G$ in eq.~(\ref{propdr}) is purely relativistic. Both the
kernel of eq.~(\ref{poten}) or the equivalent relativistic one of
Ref.~\cite{bennhold}
were used in~\cite{bennhold} in connection with the $G$ function of 
eq.~(\ref{propdr}) and the differences were found negligible. The same
comment can be made about treating the decuplet of baryons as non-relativistic in
the Lagrangians and using eq.~(\ref{propdr}) which is fully relativistic.
Expressions  for $G$ with non-relativistic propagators were used in
Ref.~\cite{angels}, and in~\cite{bennhold} calculations with both expressions
of $G$ were used and once more the results obtained were practically the same.
In the present case we do not aim at such a quantitative description as
in~\cite{bennhold}, and hence, these approximations are further justified.
 
With these differences in the two approaches it is interesting to remark that
the amplitudes obtained in the regions of interest are remarkably similar, hence
reinforcing the claim done in~\cite{lutz} about the nature of the resonances
obtained as dynamically generated. It also provides a calculational scheme which
simplifies the one done in~\cite{lutz}. The present work in fact complements the
one of~\cite{lutz} looking for poles in the complex plane to give further
support to the resonances found, investigates the $SU(3)$ structure of the
resonances found, considering the finite width of the decuplet resonances when
needed and makes a thorough study of each one of the
resonances looking also at the residues at the poles which provide partial decay
widths into the coupled channels considered. This information is necessary to
further exploit the results obtained so far and make contact with the physical
processes. One example is the use of the present information to study the
$\gamma p\rw K^+K^-p$ in the region of invariant masses of $K^-p$ above the
$\Lambda(1520)$ peak~\cite{luis}. The understanding of this background is of
primary importance in studies of $\Theta^+$ pentaquark production, as stressed by
Nakano in~\cite{nakano_penta04}.      

With $V$ given by eq.~(\ref{poten}) and $G$ by eq.~(\ref{propdr}), we solve 
the Bethe-Salpeter equation~(\ref{LS}) for coupled channels to get the
transition amplitudes $T$.

\section{Search for poles}

We look for poles of the transition matrix $T$ in the complex $\sqrt{s}$ plane.
The complex poles, 
$z_R$, appear in unphysical Riemann sheets. In the
present search we move to these sheets by changing
the sign of the complex valued momentum $q_l$ from positive to negative 
in the loop function $ G_l(z)$ of
eq.~(\ref{propdr}) for the channels which
are above threshold at an energy equal to Re($z$). This we call the second
Reimann sheet $R_2$. In the cut-off method the same can be 
achieved by using the propagator (\ref{propcutoff}) for energies below
threshold and replacing above threshold with
\be
G^{2nd}=G+2i\,\frac{p_{CM}}{\sqrt{s}}\,\frac{M}{4\pi}.
\label{defR2}
\ee
The variables on the right hand side of the above equation are
evaluated in the first (physical) 
Riemann sheet. In eq.~(\ref{defR2}), $p_{CM}$, $M$ and $\sqrt{s}$ denote the 
momentum in the centre of mass (CM), the baryon mass and the CM energy respectively. We also
define another Riemann sheet $R_3$ where we change $q_l$ from positive 
to negative 
in eq.~(\ref{propdr}) for all values of the complex energy.

Note that due to unitarity and hermitian analyticity one has the analytic
continuation of the $T$ matrix $T(\sqrt{s})=T^\ast(\sqrt{s} \ ^\ast)$, which has a
consequence that the poles corresponding to resonances appear as pairs of
complex conjugate poles. In this work we report on the pole with negative
imaginary part.
 
\begin{figure}[h]
\includegraphics[width=0.5\textwidth]{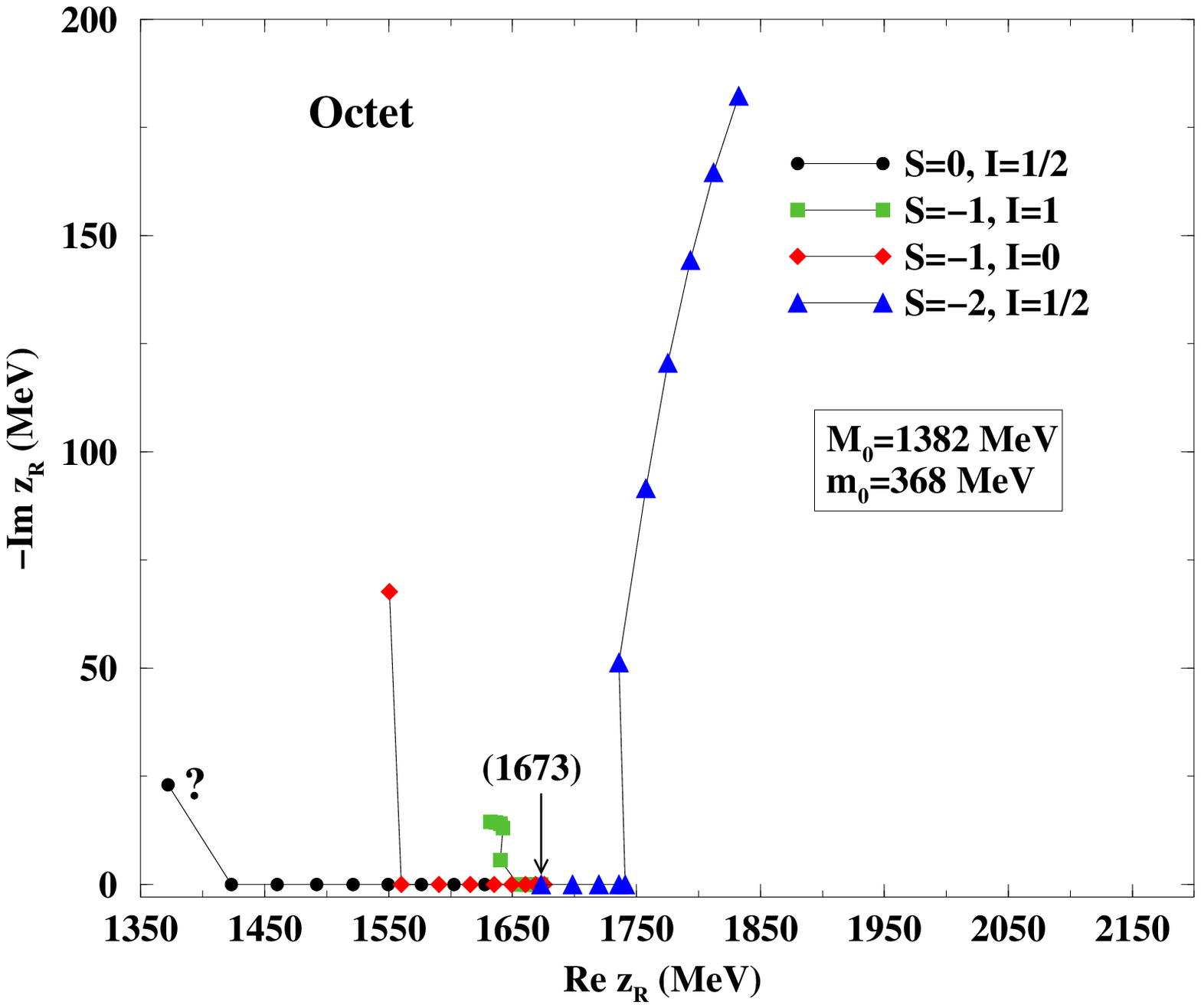}
\includegraphics[width=0.5\textwidth]{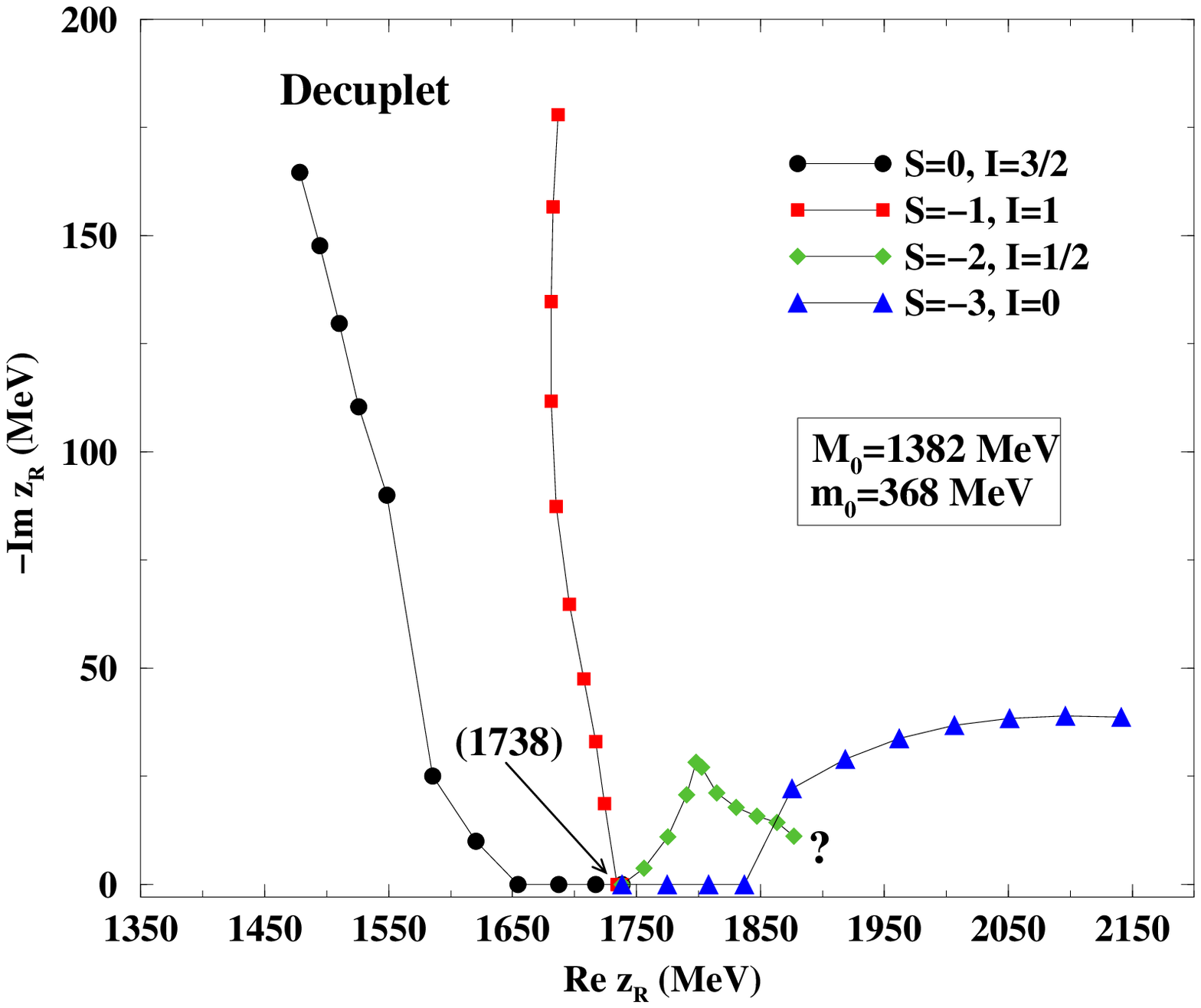}
\caption{Trajectories of the poles in the scattering amplitudes obtained by
increasing the $SU(3)$ breaking parameter $x$ from zero to one in steps of 0.1
for the octet (left) and decuplet (right) representations. The poles which
disappear are indicated by a $?$-sign.}
\label{trajfig1}
\end{figure}

We begin our search for poles in the flavour $SU(3)$ limit
which is obtained by putting common masses $M_0$ and $m_0$ respectively for the 
decuplet baryons and octet mesons.
We then study the trajectories of these poles in the
complex plane as a function of the parameter $x$ which breaks the
$SU(3)$ symmetry gradually up to the physical masses of the mesons and
baryons. The masses of the baryons and mesons are given respectively by
\begin{eqnarray}
M_i(x) &= & M_0+x(M_i-M_0),  \nonumber \\
m^{2}_{i}(x) &=& m_{0}^{2} + x (m^{2}_{i}-m^{2}_{0}), 
\end{eqnarray} 
with $0\le x \le 1$.
We use two sets of masses for the $SU(3)$ limit; a)the average mass $M_0=1382$
MeV for the decuplet and $m_0=368$ MeV for the octet mesons, and b) $M_0=1250$
MeV and $m_0=280$ MeV. 

In the $SU(3)$ limit (i.e. for $x=0$) we find two poles on
the real axis one of which is found to correspond to a bound state belonging to the 
octet and the other one to the decuplet
representation. As the symmetry is broken gradually, different branches for
each combination of $S,I$ appear. This means four branches each for the
octet and the decuplet.
We plot the resulting trajectories  
for the octet and decuplet representations in fig.~\ref{trajfig1} for parameter
set a).
For the octet (left panel),
all the trajectories coincide in the $SU(3)$ limit at 1673 MeV. Of the
four branches, all except the one with  $S=-2,\,I=\frac{1}{2}$ move to lower
energies. The one corresponding to $S=0,\,I=\frac{1}{2}$ actually 
disappears at $x=1$ where it reaches the $\De \pi$ threshold. In the case of 
the decuplet representation, all
the branches coincide at 1738 MeV in the $SU(3)$ limit. Two of the
branches move to lower energies and two shift towards higher energies
with increasing value of $x$. The pole with $S=-2
,~~I=\frac{1}{2}$ actually disappears at the $\Sg \ov K$ threshold 
in the limit of the physical masses. The poles which disappear are
marked with a $?$-sign in fig.~\ref{trajfig1}. We
then look for them in the sheet $R_3$. The results are discussed in the next section.

\begin{figure}[h]
\begin{center}
\includegraphics[width=0.5\textwidth]{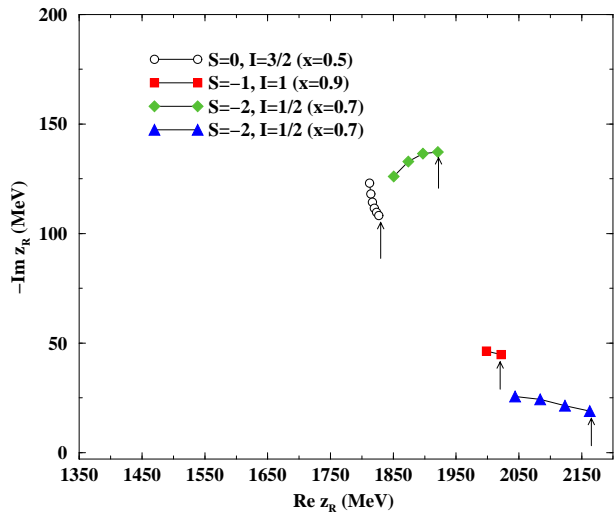}
\caption{Trajectories of some poles other than the 8 and 10 representations. The
values of $x$ given in brackets indicate the appearance of the poles and the 
arrows indicate the final position $(x=1)$ of these poles in the complex plane.}
\label{traj_oths}
\end{center}
\end{figure}
 
In addition to the poles corresponding to the octet and the decuplet which we find
already in the $SU(3)$ limit, we search for poles of the 27-plet representation in
this limit and we do not find them. However, we find that for several values of
the $SU(3)$ breaking parameter $x$, new poles in $R_2$ develop as seen in
fig.~\ref{traj_oths}. Of course we cannot associate them to the evolution of the
27-plet poles in the $SU(3)$ symmetric case (which do not show up),
but the fact that these poles are distinct from the 8 and 10 representations,
makes us conclude that they are tied to the 27-plet, but with a mixture with the
octet and decuplet, since these appear at finite values of $x$ for which the
$SU(3)$ symmetry has already been broken. There can also be a small admixture with
the 35-plet representation but since the interaction is repulsive in this case, it
should play a minimal role in the build up of the states.

\begin{figure}
\includegraphics[width=0.5\textwidth]{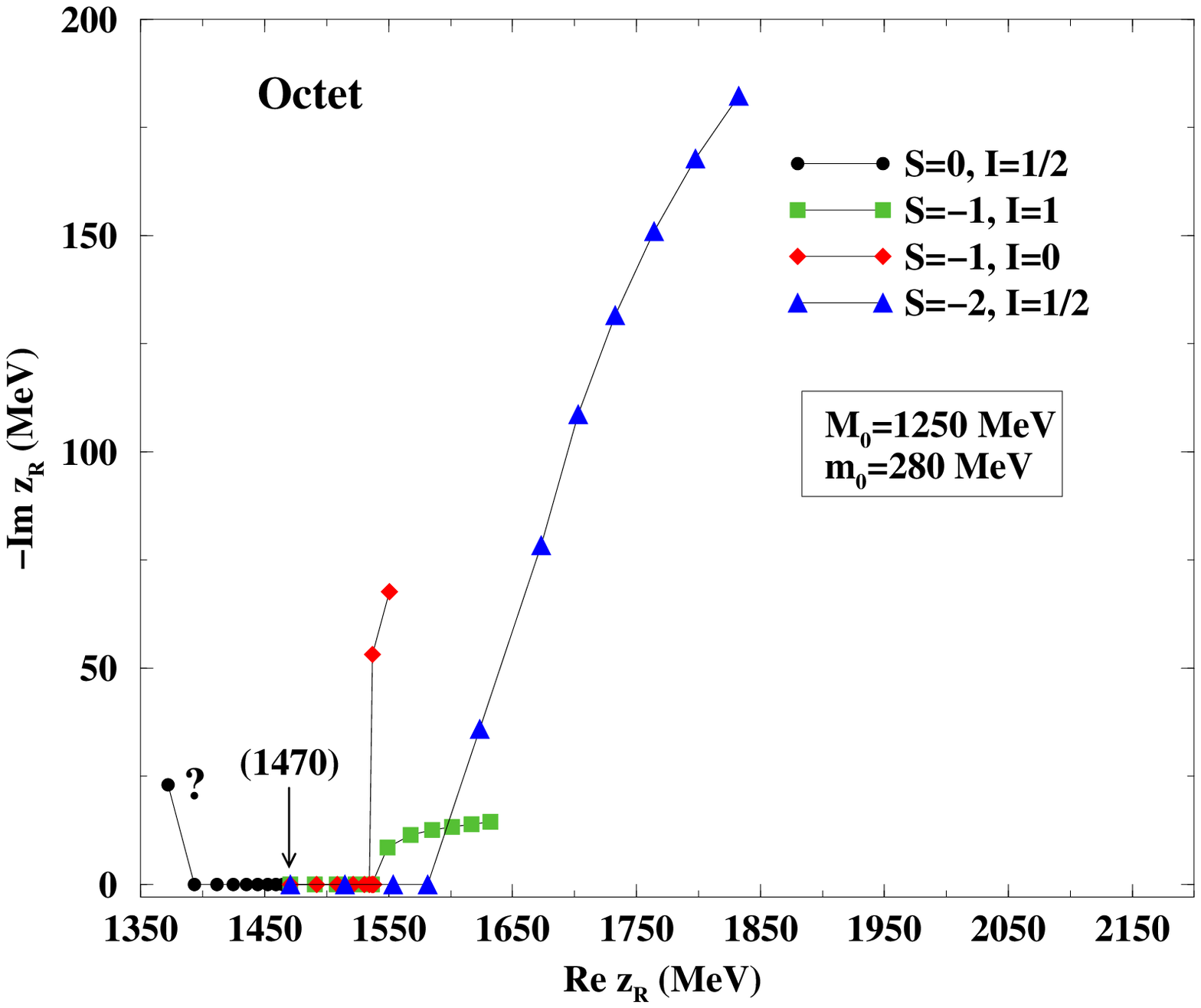}
\includegraphics[width=0.5\textwidth]{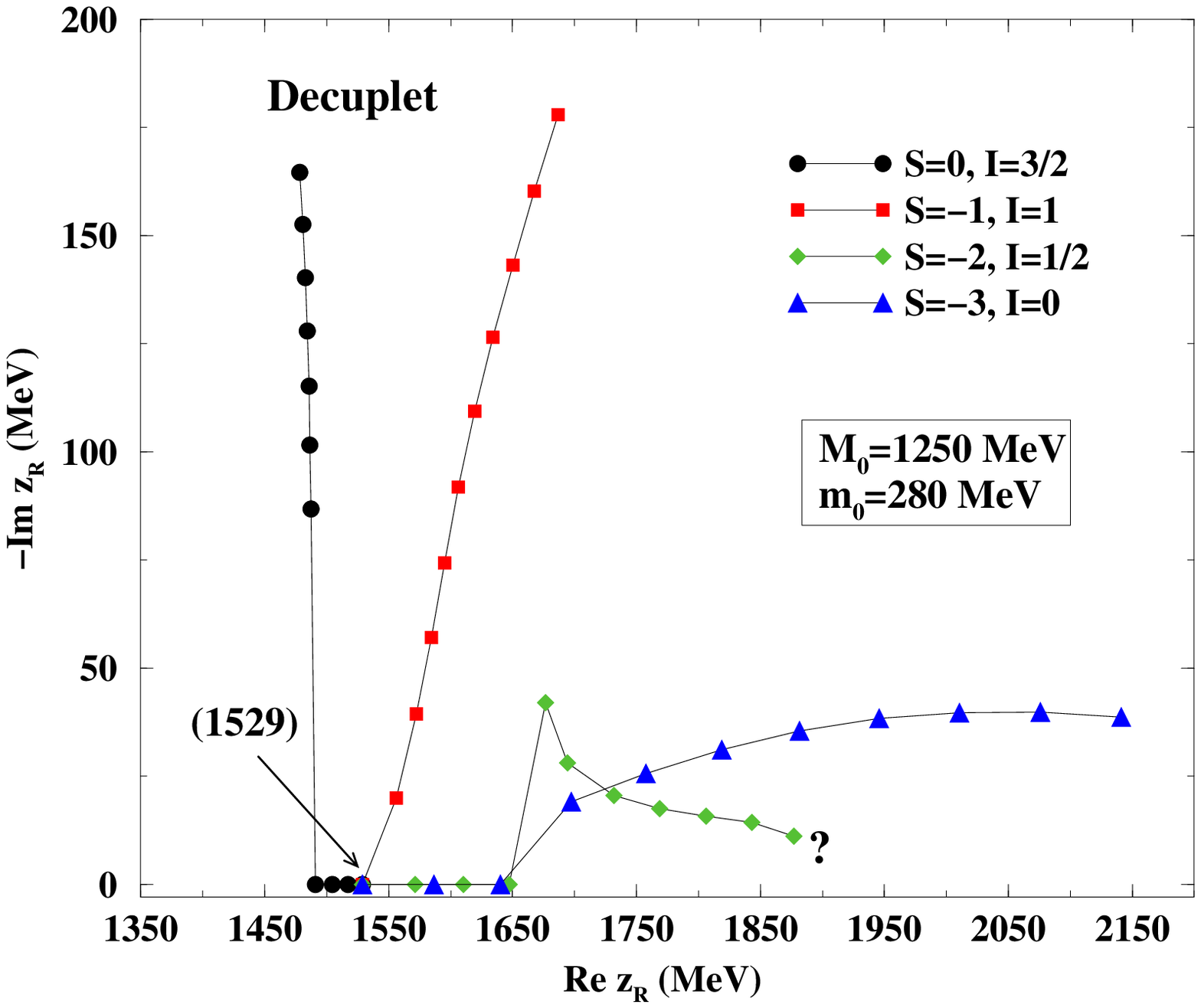}
\caption{Same as fig.~\ref{trajfig1} with $SU(3)$ limit b)}
\label{trajfig2}
\end{figure}

The trajectories of the poles belonging to the decuplet and octet obtained with
parameter set b) for the $SU(3)$ limit are shown in fig.~\ref{trajfig2}. In
this case, at $x=0$, the octet and decuplet poles appear at 1470 MeV and 1529 
MeV respectively. With increasing values of $x$, almost all of the trajectories
move towards the right to their final positions, which are the same as in
fig.~(\ref{trajfig1}) in the physical limit. It is interesting to see that the
final points of the trajectories are the same as with set a). It could have
happened that starting from a different pole in the case of $SU(3)$ symmetry,
some trajectories would have finished at a different point, revealing new poles
missed before, but this is not the case.
   
We now determine the couplings of the resonances with the different physical states 
which are the residues at the poles of the scattering matrix. 
The amplitude close to the pole at $z_R$ is identified with a form
\be
T_{ij}=\frac{g_i g_j}{z-z_R}.
\ee
We then evaluate the residues of $T_{ij}$ to get the complex valued couplings $g_i$.

The search for poles is mostly limited to what we call the second Riemann sheet $R_2$ in this
section. There is abundant literature on this
subject~\cite{polesearch1,polesearch2,polesearch3,polesearch4,polesearch5,polesearch6} and 
in some Riemann sheets one may obtain shadow poles or virtual poles. Our own
experience in this kind of problems~\cite{angels,bennhold,bennhold2,inoue,cola}
is that the second Riemann sheet is the most suited to establish the
correspondence between poles and resonances. Close to threshold the poles
corresponding to resonances might move to the sheet $R_3$ discussed in the next
section, as was found in~\cite{oller}, and hence we have investigated this
possibility in the appropriate cases. The existence of poles, together with the
amplitudes on the real axis showing the resonance structure, gives double support
to the resonances claimed. In case we did not find a pole in $R_2$ or $R_3$ and
a peak appeared at a threshold this was an indication that one had a cusp and
not a resonance and we call the attention whenever this happens.

\section{Detailed study of the ${\frac{3}{2}}^-$ resonances}

\begin{figure}[h]
\includegraphics[width=0.5\textwidth]{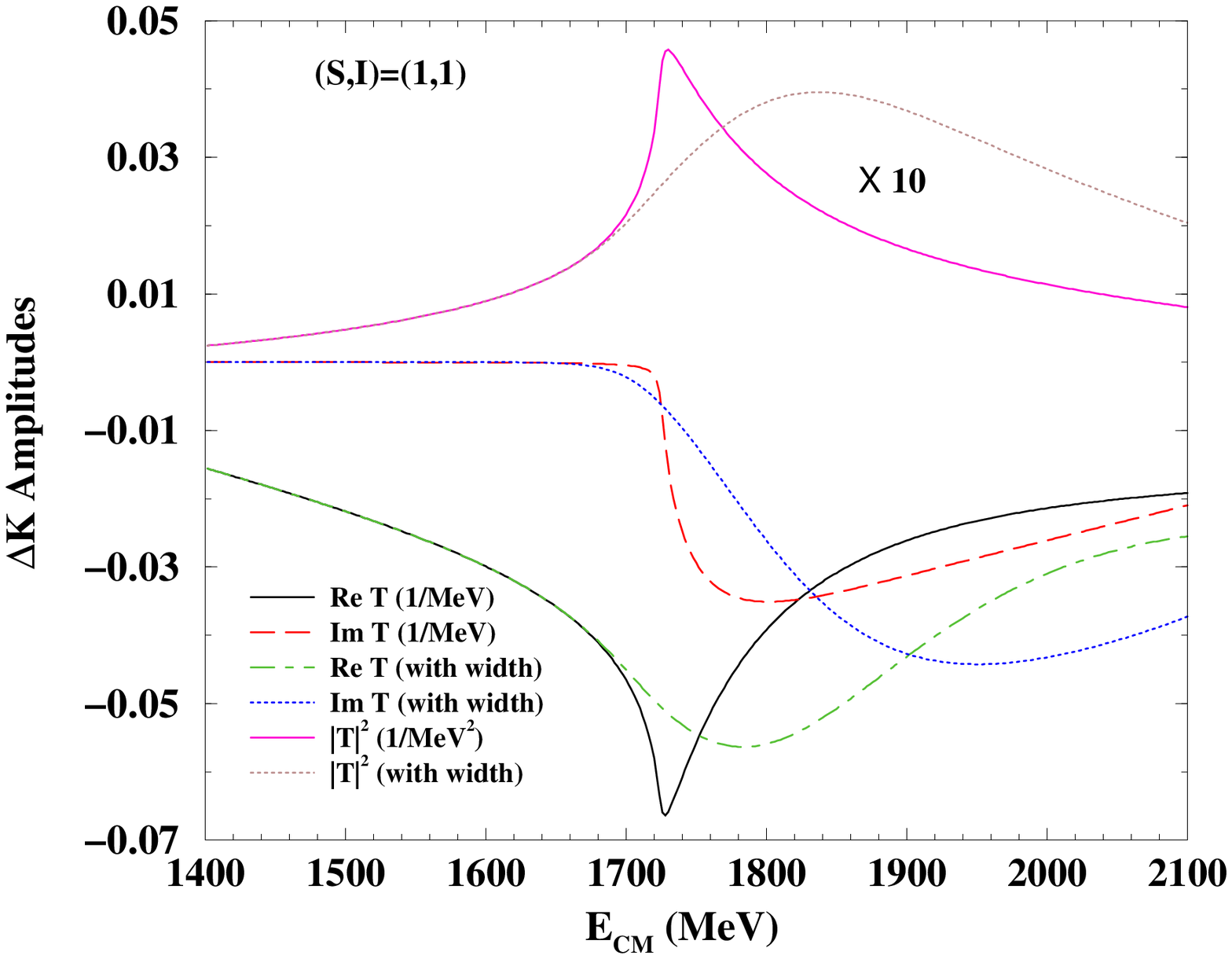}
\includegraphics[width=0.5\textwidth]{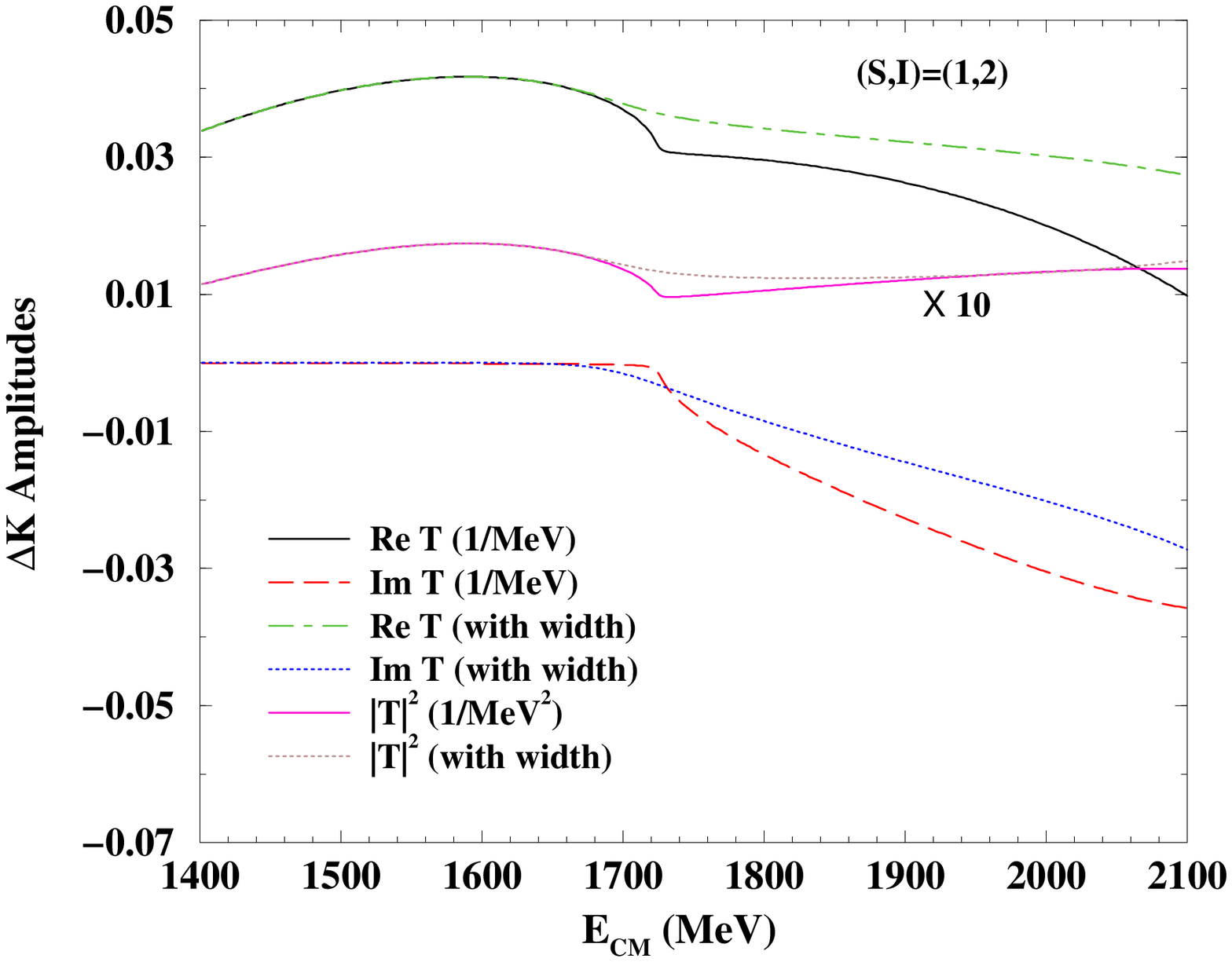}
\caption{Amplitudes for $\Delta K\rightarrow\Delta K$ for $I=1$ (left) and $I=2$ 
(right). The curves labeled 'with width' take into account
the width of the $\De$ explicitly as in~\cite{Sarkar:2004sc}.}
\label{delkfig}
\end{figure}

In the following we discuss in detail the poles of the scattering matrix in
the complex plane that we find for various values of strangeness and isospin. We
also discuss the corresponding situation as a function of the energy in the CM
frame. 

\subsection{$S=1$, $I=1$}

\begin{figure}[h]
\includegraphics[width=0.5\textwidth]{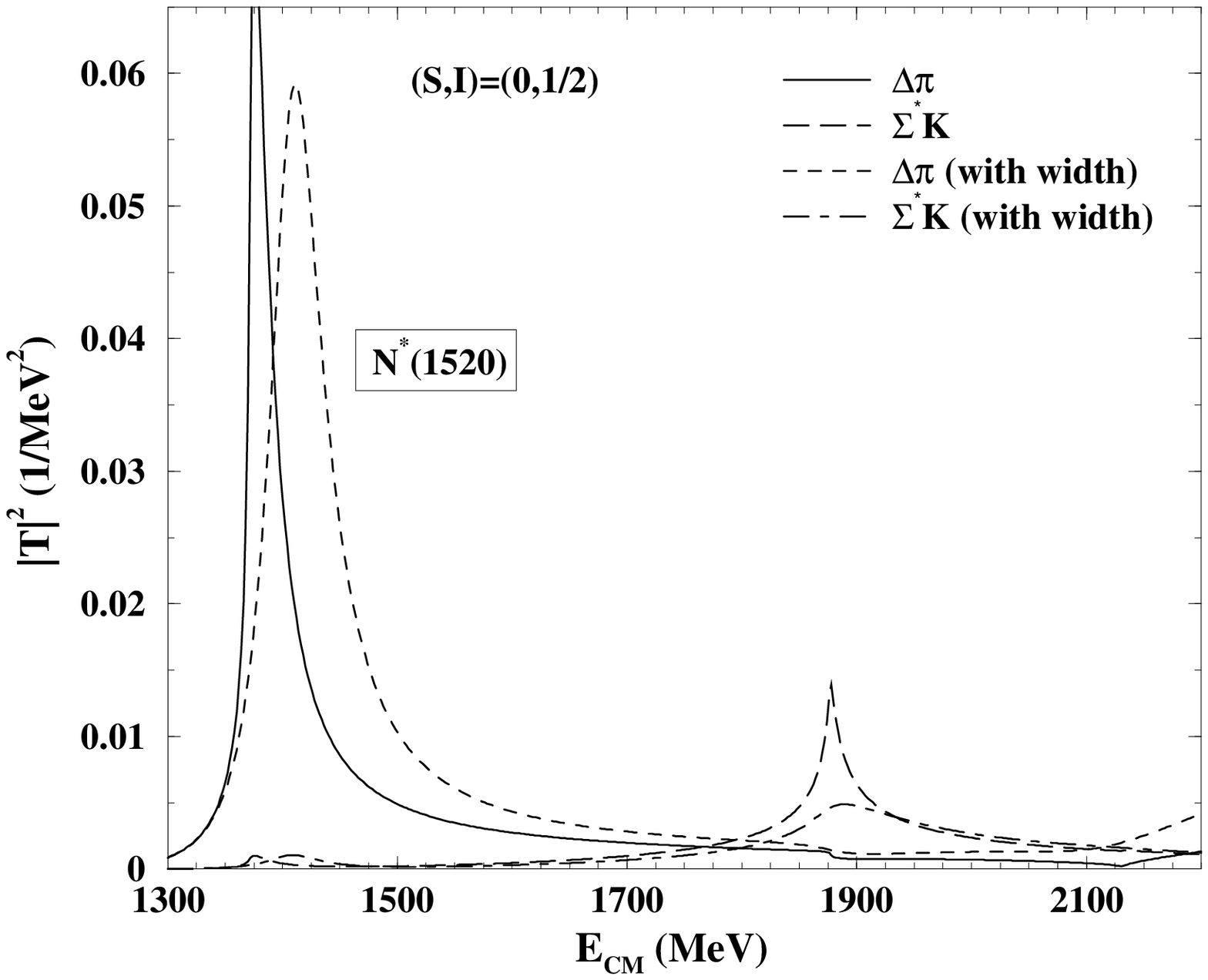}
\includegraphics[width=0.5\textwidth]{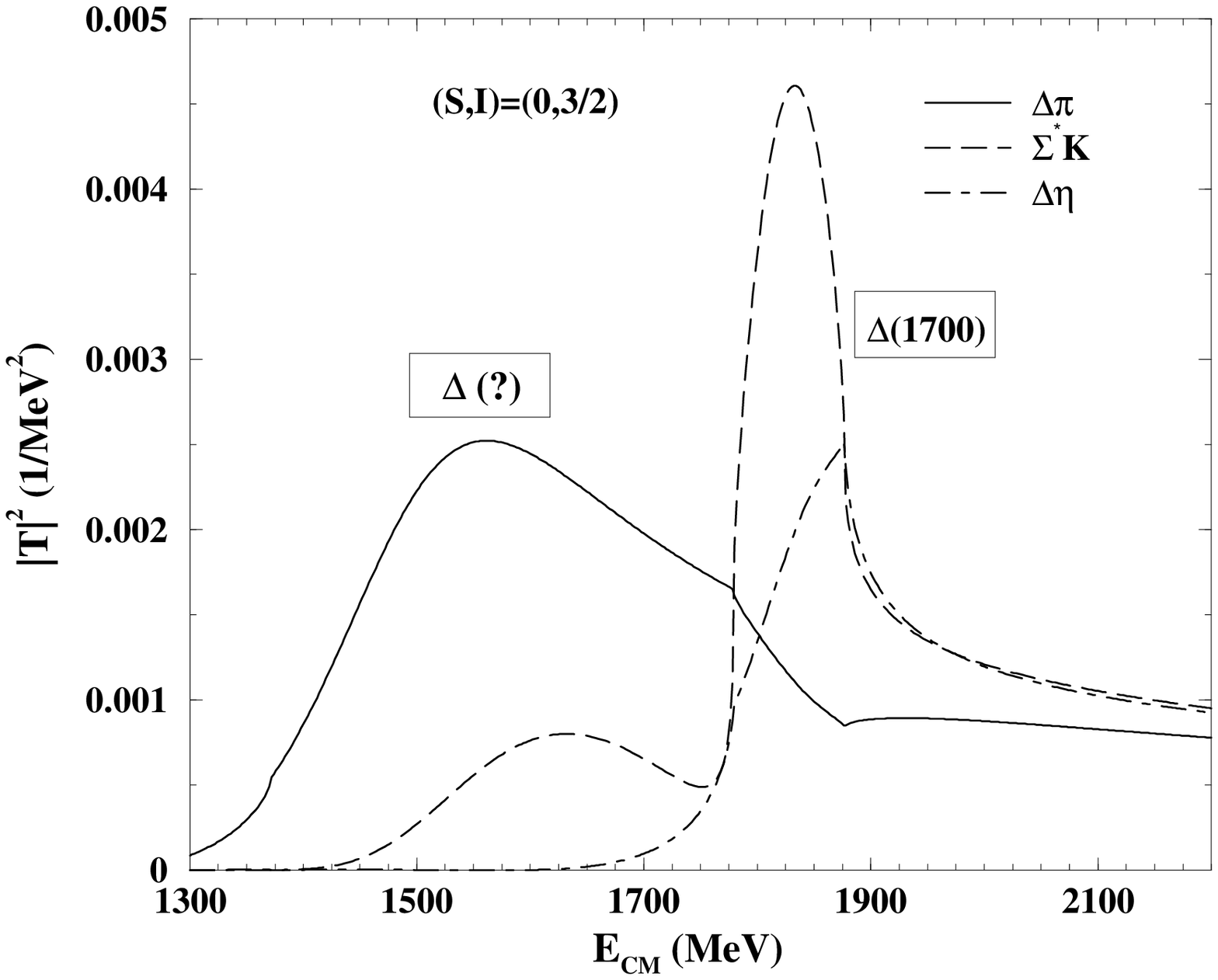}
\caption{Amplitudes for $S=0$ and $I=1/2$ (left) and $I=3/2$ (right).}
\label{fig12}
\end{figure}

The above quantum numbers belong only to the 27-plet representation. 
The interaction is attractive in the $\Delta K$ channel for $I=1$, (while the interaction
is repulsive for $I=2$ which belongs to the 35-plet representation). 
The attractive interaction leads to a large
accumulation of strength in the scattering 
amplitude close to the $\Delta K$ threshold. This is also reflected as a 
pole in the 
Riemann sheet $R_3$ at 1635 MeV on the real axis.
The singularity of the $I=1$ scattering amplitude in the Riemann sheet
$R_3$ makes the $\De K$ amplitude very different compared to the one for $I=2$
as seen in fig.~\ref{delkfig}.  
The increased strength in the $I=1$ channel accumulated close
to the $\De K$ threshold is actually a reminder of the existence of the pole in
$R_3$. For all these reasons, this singularity can
qualify as a dynamically generated resonance as claimed in~\cite{Sarkar:2004sc},
even if it does not exactly correspond to a bound state which is a pole below
threshold on the first Riemann sheet. The possibility for a resonance to emerge in
this channel was already suggested in~\cite{lutz}. 

\subsection{$S=0$, $I=\frac{1}{2}$}

In this case
the states $\De\pi$ and $\Sg K$ couple to produce two peaks 
close to the corresponding thresholds at 1372 and 1877 MeV respectively. 
This is shown in the left panel of fig.~\ref{fig12}. The
one at 1372 MeV actually corresponds to the leftmost branch of the octet
(see fig.~\ref{trajfig1}) which merges with
the $\De\pi$ threshold at $x=1$.
 We then look
in the sheet $R_3$ where we find a pole at 1328 MeV on the real axis having a 
strong coupling to the $\De\pi$ channel. 
We also evaluate the scattering amplitude incorporating the widths of the 
$\De\ (\rw N\pi)$ and the $\Sg\ (\rw \Lambda\pi)$.
This is done by adding $-i\Gamma(q^2)/2$ to the baryon energy,
$E_l(\vec q)$, in the last factor of eq.~(\ref{propcutoff}) where
\be
\Gamma (q^2)=\Gamma_0\,\frac{q_{CM}^3}{\bar q_{CM}^3}\,\,
\Theta(\sqrt{q^2}-M_N-m_\pi),
\ee
in which $q_{CM}$ and $\bar q_{CM}$ denote the momentum of the pion 
in the rest frame of the (decuplet) baryon corresponding to invariant masses
$\sqrt{q^2}$ and $M_\Delta$ (or $M_{\Sg}$) respectively~\cite{Sarkar:2004sc}. 
$\Gamma_0$ is taken as 120 MeV for the $\De$ and 35 MeV for the $\Sg$. This shifts the peak at
1372 by about 50 MeV towards higher energies as indicated by the short dashed line 
in fig.~\ref{fig12}. The closest object to this resonance which we find in the 
Particle Data Book (PDB)~\cite{Hagiwara:fs} is the $N^*(1520)$ which also couples to $N\rho$ suggesting
that this extra channel could provide a source of repulsion. Indeed the contact $\rho
N\rw\rho N$ interaction is shown to be repulsive in~\cite{Herrmann:1993za}, and
together with further corrections done in the $\rho$ interaction with the medium 
in~\cite{Cabrera:2000dx}, which can be cast into an effective $\rho N\rw\rho N$
interaction, pile up to a net $\rho N$ repulsion.

As to the peak on the threshold around 1877 MeV, it does not correspond to any pole
in the trajectories in figs.~\ref{trajfig1} and \ref{traj_oths} and neither shows up
as a pole in $R_3$. We thus attribute it to a threshold cusp. This conclusion is
corroborated by the fact that the peak does not move when we change the subtraction
constant $a$ within reasonable boundaries. 

\subsection{$S=0$, $I=\frac{3}{2}$}

In this case we have the $\De\pi$, $\Sg K$ and the $\De\eta$ states in 
coupled channels. In the second Riemann sheet we have two clear poles 
at $(1478- i165)$ and $(1827- i108)$. These correspond to a pole of the
decuplet in fig.~\ref{trajfig1} and to one of the mixed poles in 
fig.~\ref{traj_oths} respectively. In table~\ref{tcS1I3by2} we
show the values of the couplings $g_i$ to the different states for
each of these resonances. The one at 1478 MeV belongs to the
decuplet and couples mostly to the
$\De\pi$.  
 The strong coupling of the  1827 MeV resonance
to the $\Sg K$ channel would make this state qualify as a quasibound 
state of a $\Sg$ and a $K$. The modulus squared of the amplitude
as a function of the CM energy can be seen from the right panel of
fig.~\ref{fig12}.
We observe a very broad peak centred around 1550 MeV
and a pronounced peak around 1830 MeV. Note that the latter peak appears
between the two thresholds $\De\et$ (1779 MeV) and $\Sg K$ (1877 MeV). In
ref.~\cite{lutz} a clear peak also appears in
the $\De\et$ in this channel on top of the $\De\et$ threshold while a broad but small
signal is seen around 1500 MeV. Our search of poles allows us to identify the
present peaks of $|T|^2$ as dynamically generated resonances and not
threshold cusps. The pole around 1830 MeV with a width of 216 MeV
could qualify as the 4-star resonance $\De(1700)$ which has a width of 300
MeV~\cite{Hagiwara:fs}. 

There is no counterpart for the state around 1550 MeV in the PDB. However, even if
broad, it appears here distinctively as a pole in the $T$-matrix  with a clearly
visible structure in the modulus squared of the $\De\pi$ amplitude in particular. 
This is hence a case of a
missing resonance that could be searched experimentally.

\begin{table}[h]
\begin{center}
\begin{tabular}{|c|cc|cc|}
\hline
  $z_{R}$ & \multicolumn{2}{c|}{$1478 - i165$} &
\multicolumn{2}{c|}{$1827 - i108$} \\
\cline{2-5}
& $g_i$ & $|g_i|$ & $g_i$ & $|g_i|$ \\
\hline
$\De\pi$ & $2.0-i1.9$ & $2.8$ & $0.5+i0.8$ & $1.0$ \\
$\Sg K$ & $1.6-i1.6$ & $2.3$ & $3.3+i0.7$ & $3.4$ \\
$\De\eta$ & $0.3-i0.1$ & $0.3$ & $1.7-i1.4$ & $2.2$ \\
\hline
\end{tabular}
\end{center}
\caption{Couplings of the resonances with $S=0$ and $I=\frac{3}{2}$ to various
channels.}
\label{tcS1I3by2} 
\end{table}

\subsection{$S=-1$, $I=0$} 

\begin{figure}[h]
\includegraphics[width=0.5\textwidth]{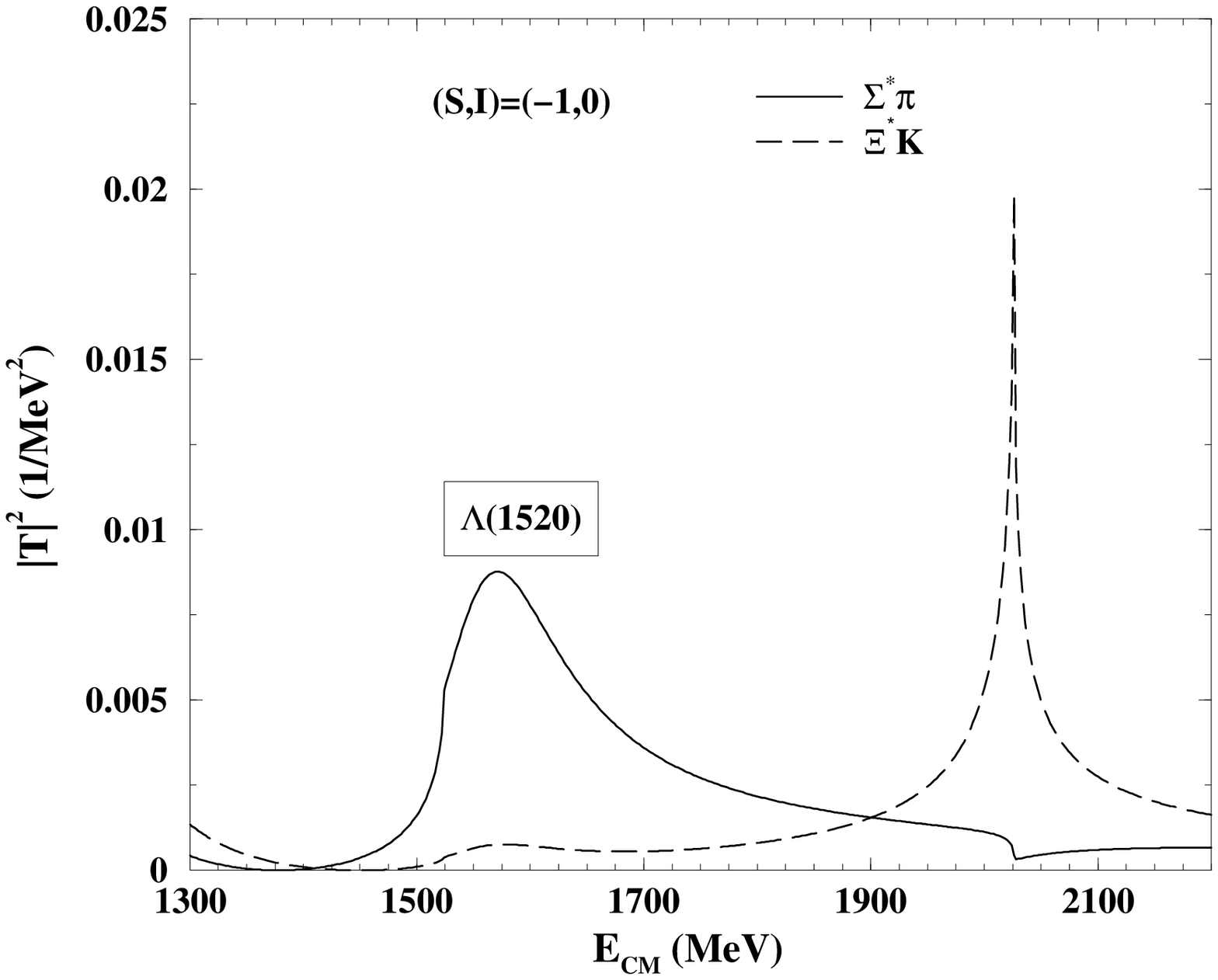}
\includegraphics[width=0.5\textwidth]{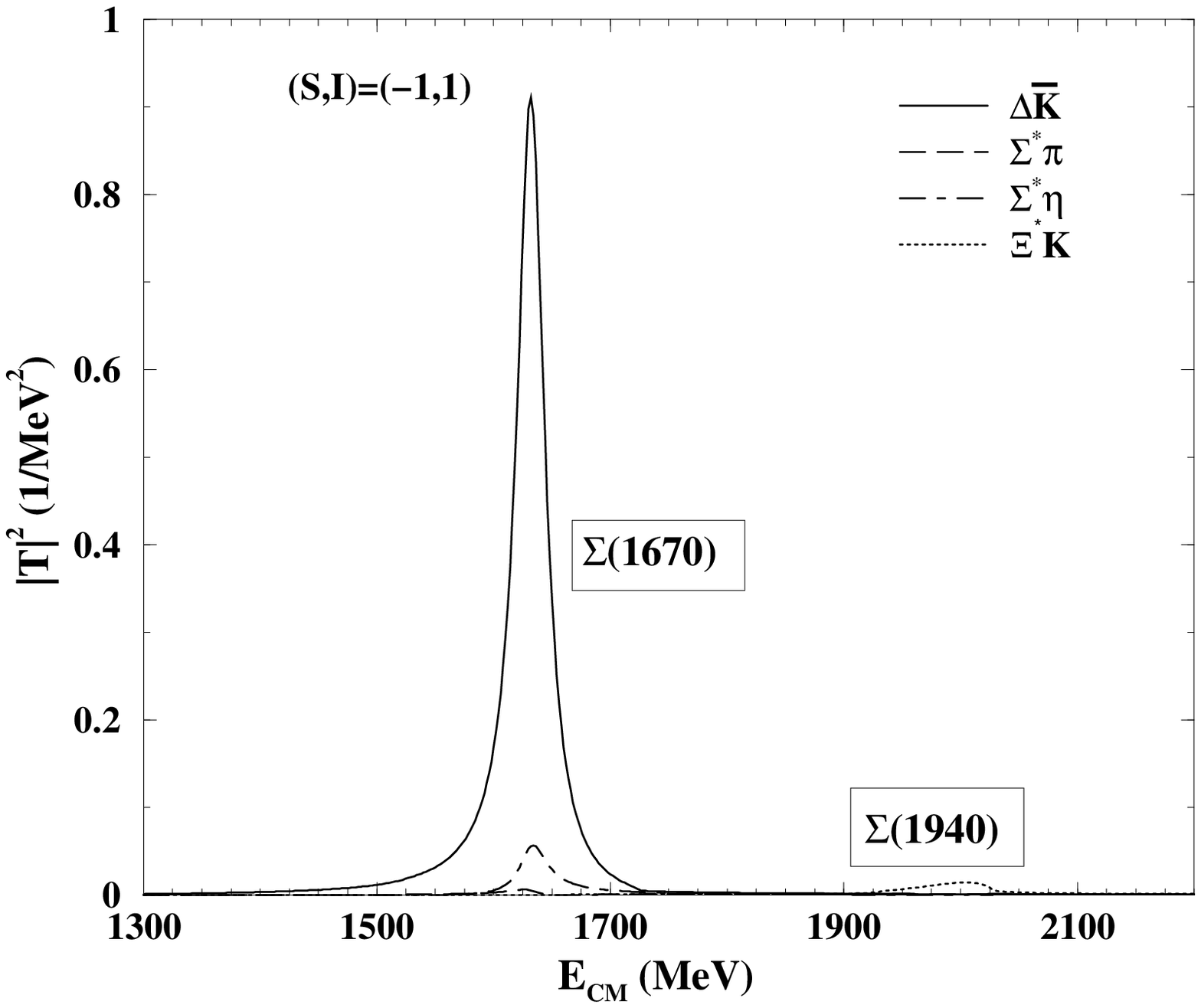}
\caption{Amplitudes for $S=-1$ and $I=0$ (left) and $I=1$ (right).}
\label{fig34}
\end{figure}

The states involved in this case are the $\Sg\pi$ and $\X K$. In the 
second Riemann sheet we find a pole at $(1550- i67)$, seen in the octet of
fig.~\ref{trajfig1}. In table~\ref{tcS-1I0} we
show the coupling of this resonance to the different states and a 
strong coupling to the $\Sg\pi$ channel is seen.
The pole
position can be made to coincide with the mass of the $\Lambda(1520)$ 
by taking the subtraction constant $a=-2.7$ in eq.~(\ref{propdr})
when it actually appears as a bound state. The absolute value of the 
coupling to $\Sg\pi$ then reduces to 1.14.
In order to do a proper comparison with the results of the PDB we then evaluate 
the width of the $\Lambda(1520)$ into $\Sg\pi$ by taking this value
of the coupling and making a convolution over the mass distribution of the
$\Sg(1385)$. One can also make an extra folding for the $\Lambda(1520)$ mass
distribution, but since its width is smaller than $\Gm_{\Sg}$ (=35 MeV), one folding is
sufficient for a rough estimate. We have
\be
\Gm_{\Sg\pi}=\int_{M_{\Sg}-\Gm_{\Sg}}^{M_{\Lambda(1520)}-m_\pi}
dM\,\frac{1}{2\pi}\ \frac{\Gm_{\Sg}}{(M-M_{\Sg})^2+(\Gm_{\Sg}/2)^2}\ 
\frac{g_{\Sg\pi}^2}{2\pi}\ q_{CM}\ \frac{M}{M_{\Lambda(1520)}}
\ee
where
\(q_{CM}=\lambda^{1/2}(M_{\Lambda(1520)}^2,M^2,m_\pi^2)/(2M_{\Lambda(1520)})\).
This gives us $\Gm_{\Sg\pi}$=2.4 MeV, which is well within the 15 MeV
width~\cite{Hagiwara:fs} of the $\Lambda(1520)$, although somewhat larger,
of the order of magnitude of the branching ratio of the $\Lambda(1520)$ to the 
$\Sg\pi$, or $\Lambda\pi\pi$ ($\Gm_i\sim 1.5$, not easily separable from the
former channel) with the inherent uncertainties of a channel so close to
threshold~\cite{Hagiwara:fs}. This is an extra test of
consistency that the association of the resonance obtained to the $\Lambda(1520)$
is reasonable, in spite of the apparent larger width obtained from the complex
pole.

The squared amplitude as a function of the CM energy
is shown in the left panel of fig.~\ref{fig34} where, 
in addition to the broad peak around 1550 MeV, we also observe a peak around 
the $\X K$ threshold i.e. 2026 MeV 
which remains static for reasonable changes of the subtraction constant. This peak around
threshold does not correspond to any pole. Indeed, the pole of the octet with
$S=-1,I=0$ corresponds to the resonance found around 1550 MeV and the decuplet does
not have a resonance with these quantum numbers (see fig.~\ref{trajfig1}). So we
should not expect another pole, except if the 27-plet representation provides it,
which is not the case as we have checked. Hence, we must conclude that this peak is
a cusp effect and, unlike in ref.~\cite{lutz}, we do not associate it
to the $\Lambda(1690)$ resonance, which however does not show up in our study.

\begin{table}[h]
\begin{center}
\begin{tabular}{|c|cc|}
\hline
  $z_{R}$ & \multicolumn{2}{c|}{$1550 - i67$} \\
\cline{2-3}
& $g_i$ & $|g_i|$ \\
\hline
$\Sg\pi$ & $2.0-i1.5$ & $2.5$ \\
$\X K$ & $0.9-i0.8$ & $1.2$ \\
\hline
\end{tabular}
\caption{Couplings of the resonance with $S=-1$ and $I=0$ to various
channels.}
\label{tcS-1I0}
\end{center}
\end{table}

\subsection{$S=-1$, $I=1$}

For these quantum numbers we observe three poles of the scattering amplitude
for the coupled channels $\De\ov K$, $\Sg \pi$, $\Sg\eta$ and $\X K$ in
the complex energy plane $R_2$ at
$(1632- i15)$, $(1687- i178)$ and $(2021- i45)$. The first pole appears in
the evolution of the octet poles and the second one in the decuplet. The third pole
is one of those found in fig.~\ref{traj_oths} which we tied to the 27-plet
representation as discussed above. The couplings of the resonances to the different
states are shown in table~\ref{tcS-1I1}. The pole at 1632 MeV is reflected
as a peak on the real axis with a very strong coupling to $\De\kb$ as seen in
fig.~\ref{fig34} (right). Its
position is found to be very sensitive to the value of the subtraction constant.
This peak can be well associated to with the 4-star resonance $\Sigma(1670)$ which
has a width of 60 MeV. The width of about 30 MeV found from the complex pole
position would go mostly to the $\Sg\pi$ since the other channels are closed.
Although no numbers are provided in the PDB, this decay is listed there and we get
from this channel a smaller width than the total, which gets at least half of the
strength from the meson-baryon decay channels. It becomes clear that further
experimental research into the meson-meson-baryon decay channels would shed light
on these theoretical ideas exploited here. 

Our approach develops a distinct second pole at a nearby position (1687 MeV)
but with a much larger width which does not allow it to show up in the $|T|^2$ plot
in the real axis of fig.~\ref{fig34}. The third pole which couples strongly to
the $\X K$ produces a small bump on the real axis and could well
correspond to the $\Sigma(1940)$ (right of fig.~\ref{fig34}) as also claimed
in~\cite{lutz}. The width of 90 MeV for this resonance found here is
also reasonably smaller than the total width of around 220 MeV mentioned in the
PDB, which partly goes to the meson-baryon decays. Once more, the experimental
study of the meson-meson-baryon channels would be more useful as a test of the
present theory.

\begin{table}[h]
\begin{center}
\begin{tabular}{|c|cc|cc|cc|}
\hline
  $z_{R}$ & \multicolumn{2}{c|}{$1632 - i15$} &
\multicolumn{2}{c|}{$1687 - i178$} & \multicolumn{2}{c|}{$2021 - i45$}  \\
\cline{2-7}
& $g_i$ & $|g_i|$ & $g_i$ & $|g_i|$ & $g_i$ & $|g_i|$ \\
\hline
$\De\kb$ & $3.7-i0.03$ & $3.7$ & $0.4-i1.7$ & $1.8$ & $0.4-i0.5$ & $0.6$ \\
$\Sg \pi$ & $1.1+i0.4$ & $1.1$ & $2.2-i2.0$ & $3.0$ & $0.3+i0.8$ & $0.8$ \\
$\Sg\eta$ & $1.8-i0.3$ & $1.9$ & $1.9+i0.6$ & $1.9$ & $1.0-i0.7$ & $1.2$ \\
$\X K$ & $0.3+i0.5$ & $0.6$ & $2.7-i1.4$ & $3.0$ & $2.5+i1.0$ & $2.7$ \\
\hline
\end{tabular}
\caption{Couplings of the resonances with $S=-1$ and $I=1$ to various
channels.}
\label{tcS-1I1}
\end{center}
\end{table}

\subsection{$S=-2$, $I=\frac{1}{2}$}
 
Here there are the four states $\Sg \ov K$, $\X\pi$, $\X\eta$ and 
$\Om K$ which couple to each other in the scattering amplitude.
On the real axis (see the left of fig.~\ref{fig56}) we observe a peak structure near the $\Sg \ov K$
threshold (1877 MeV). In the complex plane $R_2$, this corresponds
to one of the four branches of the decuplet which disappears at the
$\Sg \ov K$ threshold for $x=1$ (see right panel of fig.~\ref{trajfig1}). 
We then look in $R_3$
where we find a pole on the real axis at 1660 MeV.  
Its position shifts with changes in the 
subtraction constant. We infer that it is indeed the 4-star resonance $\Xi(1820)$ which
has a width of about $24^{+15}_{-10}$ MeV. It is worth mentioning that at $x=0.9$ this resonance
still shows up as a pole in $R_2$ at $1863- i14$ which means a width of 28 MeV.
At this energy the most relevant channel for decay is the $\X\pi$. The width of
28 MeV would correspond practically to this channel but on the real axis,
because of the Flatt\'{e} effect~\cite{flatte}, with the opening of the $\Sg \ov K$ channel to
which the resonance couples strongly, the apparent width is smaller, close to 18
MeV, as one can see in fig.~\ref{fig56} (left panel). This would be compatible
with the 30$\pm 15\%
$ branching of the $\Xi(1820)$ into the $\X\pi$. We have also done fine tuning
to bring the pole position to lower energies, however, the proximity of the
second pole at $1832-i182$ MeV to the nominal energy of the $\Xi(1820)$ distorts
the other pole when we try to bring it closer to 1820 MeV. So, no new valuable
information can be obtained with this exercise.

We also observe another smaller and relatively narrow peak on the real axis at
2162 MeV which is just below the $\Om K$ threshold (2165 MeV) in
the left of fig.~\ref{fig56}. It is
found to couple very strongly to the $\Om K$ (see table~\ref{tcS-2I1by2}). This
strong coupling to the $\Om K$ compared to the other much smaller couplings and the
fact that the peak appears below the $\Om K$ threshold makes this state qualify
clearly as a quasibound $\Om K$ state. There are several candidates of the $\Xi$
resonance in the vicinity of this mass with spin and parity unknown. It would be
interesting to confirm experimentally the quantum numbers of these resonances to
allow us the identification of the resonance found. Alternatively, one can make
progress by measuring the partial decay widths which we can predict in this model.
This could also help identify our resonance with a particular one and indirectly
predict spin and parity of this resonance as was done in~\cite{bennhold2} to predict the
spin and parity of the $\Xi(1620)$.

In table~\ref{tcS-2I1by2} we see two more states. The first one at 1832 MeV
corresponds to the one found in the evolution of the octet (see
fig.~\ref{trajfig1}). The second one at 1920 MeV (as well as the one at 2162 MeV
mentioned above) are those found in fig.~\ref{traj_oths} corresponding to some
mixture with the 27-plet representation. These two states, however, are very
broad and they do not show up in the $|T|^2$ plot of the amplitudes in the real
plane (fig.~\ref{fig56}), hence the chances of observation are not too bright.  

\begin{table}[h]
\begin{center}
\begin{tabular}{|c|cc|cc|cc|cc|}
\hline
  $z_{R}$ & \multicolumn{2}{c|}{$1863 - i14(x=0.9)$} & \multicolumn{2}{c|}{$1832
  - i182$} &
\multicolumn{2}{c|}{$1920 - i137$} & \multicolumn{2}{c|}{$2162 - i19$}  \\
\cline{2-9}
& $g_i$ & $|g_i|$ & $g_i$ & $|g_i|$ & $g_i$ & $|g_i|$ & $g_i$ & $|g_i|$ \\
\hline
$\Sg\kb$ & $1.9+i0.7$ & $2.0$ & $1.8-i1.1$ & $2.1$ & $1.1+i0.1$ & $1.1$ & $0.3-i0.4$ & $0.5$ \\
$\X \pi$ & $0.5+i0.9$ & $1.1$ & $2.3-i1.8$ & $2.9$ & $1.1-i1.7$ & $2.0$ & $0.2+i0.7$ & $0.7$ \\
$\X\eta$ & $2.5+i0.2$ & $2.6$ & $1.4+i1.3$ & $1.9$ & $3.5+i1.7$ & $3.8$ & $0.4-i0.3$ & $0.5$ \\
$\Om K$ & $0.1-i0.7$ & $0.7$ & $2.3-i0.9$ & $2.4$ & $1.6-i0.4$ & $1.7$ & $2.1+i0.9$ & $2.3$ \\
\hline
\end{tabular}
\caption{Couplings of the resonances with $S=-2$ and $I=\frac{1}{2}$ to various
channels. Note that the couplings for the 1877 MeV resonance are evaluated at
$x=0.9$.}
\label{tcS-2I1by2}
\end{center}
\end{table}

\subsection{$S=-3$, $I=0$}

\begin{figure}[h]
\includegraphics[width=0.5\textwidth]{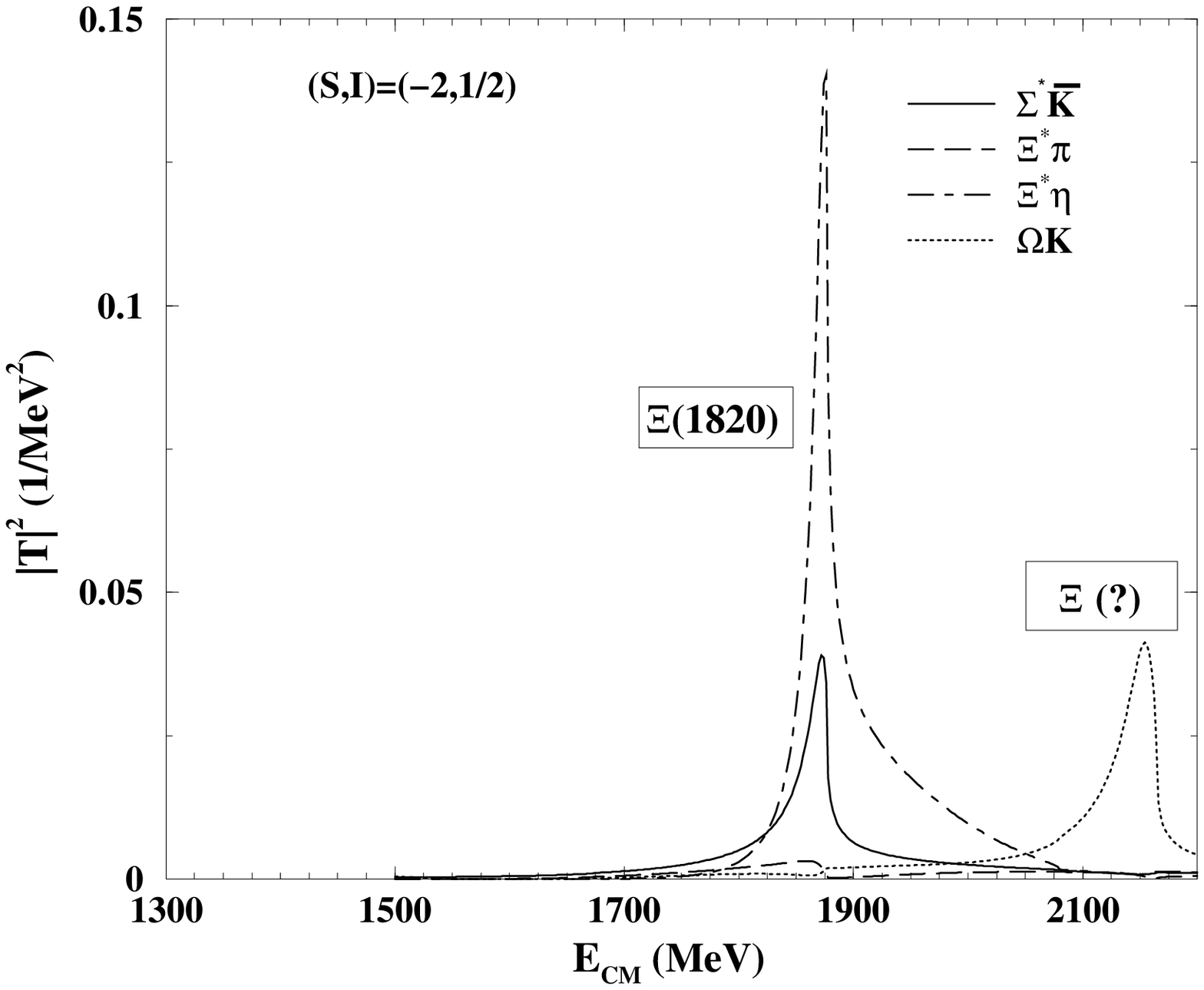}
\includegraphics[width=0.5\textwidth]{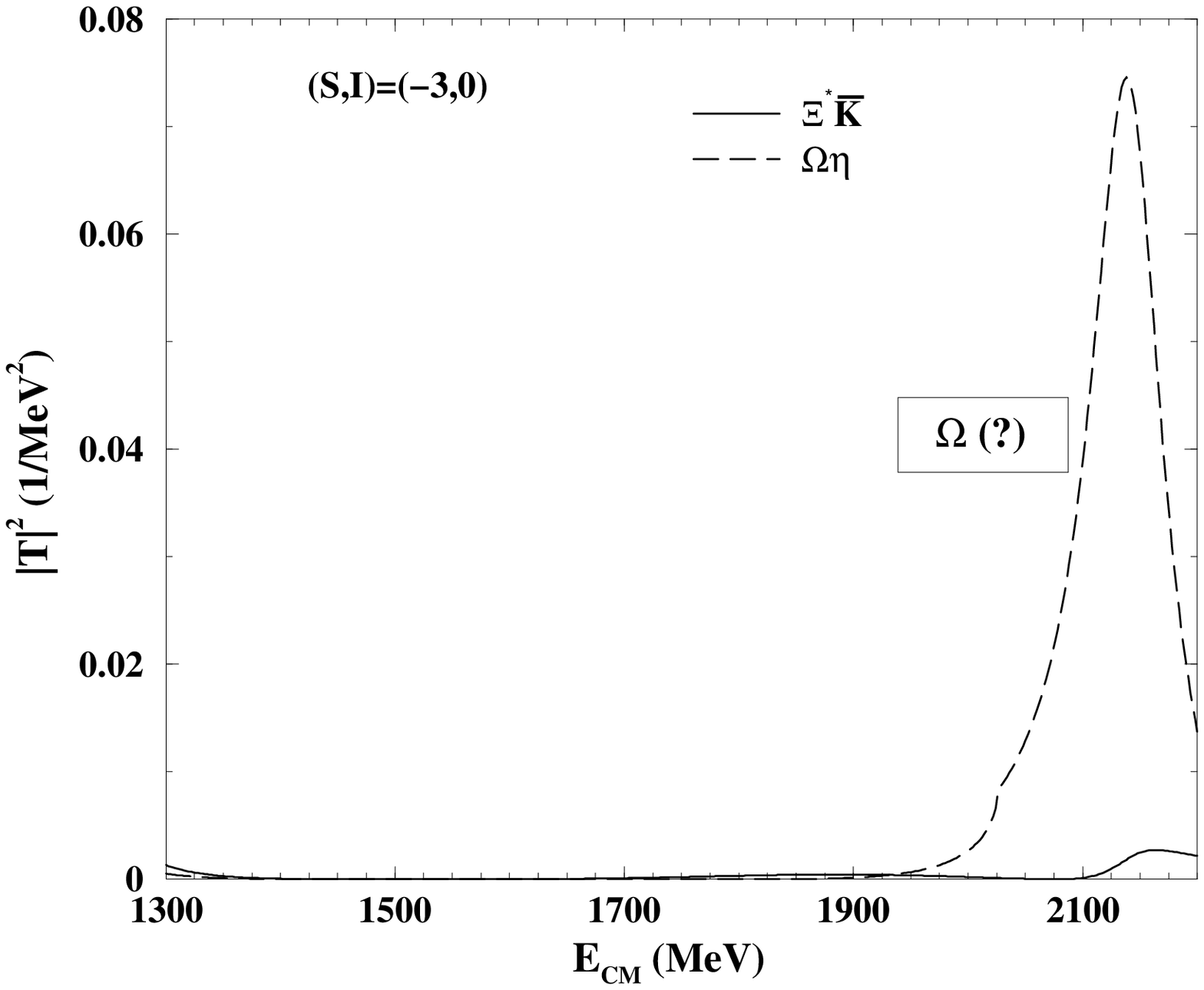}
\caption{Amplitudes for $S=-2$ and $I=1/2$ (left) and $S=-3$, $I=0$ (right).}
\label{fig56}
\end{figure}

There are two coupled states in this case, the $\X\ov K$ and the
$\Om\eta$. We find a pole at $(2141- i38)$ in the complex plane $R_2$
which is reflected as a peak in the amplitude on the real
axis at 2141 MeV (right panel of fig.~\ref{fig56}). The position shifts with the value of the subtraction
constant and has a strong coupling to the $\Om\eta$ channel as seen in
table~\ref{tcS-3I0}.
There is only one excited resonance with these quantum numbers
tabulated in the PDB~\cite{Hagiwara:fs}, 
the $\Om(2250)$ of which neither the spin nor the parity are known. The
identification of our resonance with this one is possible, given the fact that
the decay mode $\X\kb$ predicted here is one of those quoted as seen in the PDB.
The partial decay width that we obtain for that channel is about 34 MeV, which
is reasonably smaller than the $55\pm 18$ MeV total width of the $\Om(2250)$
tabulated in the PDB. Once more, the thorough investigation of these decay
channels will shed light on the predictions done here.
 
\begin{table}[h]
\begin{center}
\begin{tabular}{|c|cc|}
\hline
  $z_{R}$ & \multicolumn{2}{c|}{$2141 - i38$} \\
\cline{2-3}
& $g_i$ & $|g_i|$ \\
\hline
$\X\kb$ & $1.1-i0.8$ & $1.4$ \\
$\Om\et$ & $3.3+i0.4$ & $3.4$ \\
\hline
\end{tabular}
\caption{Couplings of the resonance with $S=-3$ and $I=0$ to various
channels.}
\label{tcS-3I0}
\end{center}
\end{table} 

\section{Conclusions}

  We have done a systematic study of the interaction of the baryon decuplet
with the meson octet taking the dominant lowest order chiral Lagrangian, which
accounts for the Weinberg Tomozawa term. We have looked in detail at the 
resonances which are generated dynamically by this interaction by searching for
poles in the complex plane in different Riemann sheets. The search was done
systematically, starting from an $SU(3)$ symmetric situation where the masses of
the baryons are made equal and the same is done with the masses of the mesons. In this
case we found attraction in an octet, a decuplet and the 27 representation,
while the interaction was repulsive in the 35 representation.  In the $SU(3)$
symmetric case all states of the $SU(3)$ multiplet are degenerate and the
resonances appear as bound states with no width. As we gradually break $SU(3)$
symmetry by changing the masses, the degeneracy is broken and the states with
different strangeness and isospin split apart generating pole trajectories in the
complex plane which
lead us to the  physical situation in the final point.  This systematic search
allows us to trace the poles to their $SU(3)$ symmetric origin, although there is
 a mixing of representations when the symmetry is broken.   In addition we also
 find poles which only appear for a certain amount of symmetry breaking and thus
have no analog in the symmetric case. 

  We evaluated the residues of the poles from where the couplings of the
resonances to the different coupled channels was found and this allowed us to
make predictions for partial decay widths into a decuplet baryon  and a meson.
There is very limited experimental information on these decay channels but,
even then, it represents an extra check of consistency of the results  which
allowed us to more easily identify the resonances found with some resonances
known, or state that the resonance should correspond to a new resonance not yet
reported in the Particle Data Book. In particular, in view of the information of
the pole positions and couplings to channels we could associate some of the
resonances found to the $N^*(1520)$, $\De(1700)$, $\Lambda(1520)$, $\Sigma(1670)$,
$\Sigma(1940)$, $\Xi(1820)$ resonances tabulated in the PDB, in agreement with the
findings of~\cite{lutz} based on peaks of the speed plot. However, a peak at the
$\X K$ threshold (2026 MeV) suggested to correspond to the $\Lambda(1690)$
in~\cite{lutz} does not show up here as a pole, but only as a cusp effect,
insensitive to the variations of the input parameters. 

We could also favour the
correspondence of a resonance found in $S=-3,\ I=0$ to the $\Om(2250)$ on the basis
of the quantum numbers, position and compatibility of the partial decay width found
with the total width of the $\Om(2250)$.

We also found several extra resonances, well identified by poles in the complex
plane which do not have a correspondent one in the PDB. Some of them are too 
broad,
which could justify the difficulty in their observation, but two other resonances,
the $\De(1500)$ with a width around 300 MeV and the $\Xi(2160)$ with a width of
about 40 MeV stand much better chances of observation. The first one because of its
large strength in the $\De\pi$ channel and the second one because of its
narrowness. 

In addition, our study produces couplings of the resonances to baryon-meson
channels which could facilitate the identification when further information on
these branching ratios is available.

Another output of our results is that in some cases the resonances found couple
very strongly to some channels with the threshold above the resonance energy,
what makes them qualify as approximately single channel quasibound meson-baryon
states.

The experimental implications of this work are important since clear predictions
on the partial decay widths into a baryon of the $\De$ decuplet and a meson of
the pion octet are made, which are amenable to experimental observation.

Another case was done about the $\De K$ resonance found in $S=1,\ I=1$ which
shows up as a pole in $R_3$ and leads to $\De K$ cross sections much larger than
the corresponding $\De K$ cross sections in $I=2$. These two cross sections
could be seen in $\De K$ production in $pp\rw\Lambda\De K$ and $pp\rw\Sigma\De
K$ reactions which could provide evidence for this new 'exotic pentaquark'
state, although its structure is more efficiently taken into account in the
meson baryon picture where it has been generated.

\section{Acknowledgements}

We would like to thank D. Jido for checking the $SU(3)$ matrix elements.
One of us (S.Sarkar) wishes to acknowledge support from the Ministerio de
Educacion y Ciencia on his stay in the program of doctores y tecnologos
extranjeros. 
This work is partly supported by DGICYT contract number BFM2003-00856,
and the E.U. EURIDICE network contract no. HPRN-CT-2002-00311.

\newpage

\section*{Appendix I}
\setcounter{section}{0}
\renewcommand{\thesection}{\arabic{section}}

In the following we tabulate the $C_{ij}$ coefficients for all possible
reactions obtained from eq.~(\ref{lag2}) for various values of strangeness and charge.

\section{\ul{$S=1$}}
\subsection{$Q=-1$}
$\Dm\kz\rw\Dm\kz~~~~C=-3$ 

\subsection{$Q=0$}
\begin{tabular}{c|cc}
& $\Dm\kp$ & $\Dz\kz$\\
\hline
$\Dm\kp$ & $0$ & $-\sqrt{3}$\\
$\Dz\kz$ & & $-2$\\
\end{tabular}

\subsection{$Q=1$}
\begin{tabular}{c|cc}
& $\Dz\kp$ & $\Dp\kz$\\
\hline
$\Dz\kp$ & $-1$ & $-2$ \\
$\Dp\kz$ & & $-1$\\
\end{tabular}

\subsection{$Q=2$}
\begin{tabular}{c|cc}
& $\Dpp\kz$ & $\Dp\kp$\\
\hline
$\Dpp\kz$ & $0$ & $-\sqrt{3}$ \\
$\Dp\kp$ & & $-2$\\
\end{tabular}

\subsection{$Q=3$}
$\Dpp\kp\rw\Dpp\kp~~~~C=-3$ 

\section{\ul{$S=0$}}
\subsection{$Q=-2$}
$\Dm\pim\rw\Dm\pim~~~~C=-3$

\subsection{$Q=-1$}
\begin{tabular}{c|cccc}
& $\Dm\piz$ & $\Dm\et$ & $\Dz\pim$ & $\Sm\kz$\\
\hline
$\Dm\piz$ & $0$ & $0$ & $-\sqrt{6}$ & $-\sq{\fr{3}{2}}$\\
$\Dm\et$ & & $0$ & $0$ & $\fr{3}{\sq{2}}$\\
$\Dz\pim$ & & & $-1$ & $1$\\
$\Sm\kz$ & & & & $-1$\\
\end{tabular}

\subsection{$Q=0$}
\begin{tabular}{c|cccccc}
& $\Dm\pip$ & $\Dz\piz$ & $\Dz\et$ & $\Dp\pim$ & $\Sm\kp$ & $\Sz\kz$\\
\hline
$\Dm\pip$ & $3$ & $\sqrt{6}$ & $0$ & $0$ & $\sqrt{3}$ & $0$\\
$\Dz\piz$ & & $0$ & $0$ & $-2\sq{2}$ & $\sq{\fr{1}{2}}$ & $-1$\\
$\Dz\et$ & & & $0$ & $0$ & $\sq{\frac{3}{2}}$ & $\sq{3}$\\
$\Dp\pim$ & & & & $1$ & $0$ & $\sq{2}$\\
$\Sm\kp$ & & & & & $1$ & $-\sq{2}$\\
$\Sz\kz$ & & & & & & $0$\\
\end{tabular} 

\subsection{$Q=1$}
\begin{tabular}{c|cccccc}
& $\Dz\pip$ & $\Dp\piz$ & $\Dp\et$ & $\Dpp\pim$ & $\Sz\kp$ & $\Sp\kz$\\
\hline
$\Dz\pip$ & $1$ & $2\sq{2}$ & $0$ & $0$ & $\sq{2}$ & $0$\\
$\Dp\piz$ & & $0$ & $0$ & $-\sq{6}$ & $1$ & $-\sq{\fr{1}{2}}$\\
$\Dp\et$ & & & $0$ & $0$ & $\sq{3}$ & $\sq{\frac{3}{2}}$\\
$\Dpp\pim$ & & & & $1$ & $0$ & $\sq{3}$\\
$\Sz\kp$ & & & & & $0$ & $-\sq{2}$\\
$\Sp\kz$ & & & & & & $1$\\
\end{tabular}

\subsection{$Q=2$}
\begin{tabular}{c|cccc}
& $\Dp\pip$ & $\Dpp\piz$ & $\Dpp\et$ & $\Sp\kp$\\
\hline
$\Dp\pip$ & $-1$ & $\sq{6}$ & $0$ & $1$\\
$\Dpp\piz$ & & $0$ & $0$ & $\sq{\frac{3}{2}}$\\
$\Dpp\et$ & & & $0$ & $\fr{3}{\sq{2}}$ \\
$\Sp\kp$ & & & & $-1$\\
\end{tabular} 

\subsection{$Q=3$}
$\Dpp\pip\rw\Dpp\pip~~~~C=-3$

\section{\ul{$S=-1$}}
\subsection{$Q=-2$}
\begin{tabular}{c|ccc}
& $\Dm\km$ & $\Sm\pim$\\ 
\hline
$\Dm\km$ & $0$ & $-\sq{3}$\\
$\Sm\pim$ & & $-2$\\
\end{tabular}

\subsection{$Q=-1$}
\begin{tabular}{c|cccccc}
& $\Dm\kbz$ & $\Dz\km$ & $\Sm\piz$ & $\Sm\et$ & $\Sz\pim$ & $\Xm\kz$\\
\hline
$\Dm\kbz$ & $3$ & $\sq{3}$ & $\sq{\frac{3}{2}}$ & $-\fr{3}{\sq{2}}$ & $0$ &
$0$\\
$\Dz\km$ & & $1$ & $-\sq{\fr{1}{2}}$ & $-\sq{\frac{3}{2}}$ & $-\sq{2}$ & $0$ \\
$\Sm\piz$ & & & $0$ & $0$ & $-2$ & $-\sq{2}$\\
$\Sm\et$ & & & & $0$ & $0$ & $\sq{6}$\\
$\Sz\pim$ & & & & & $0$ & $\sq{2}$\\
$\Xm\kz$ & & & & & & $1$\\
\end{tabular}

\subsection{$Q=0$}
\begin{tabular}{c|cccccccc}
& $\Dz\kbz$ & $\Dp\km$ & $\Sm\pip$ & $\Sz\piz$ & $\Sz\et$ & $\Sp\pim$ & $\Xm\kp$
& $\Xz\kz$\\
\hline
$\Dz\kbz$ & $2$ & $2$ & $-1$ & $1$ & $-\sq{3}$ & $0$ & $0$ & $0$\\
$\Dp\km$ & & $2$ & $0$ & $-1$ & $-\sq{3}$ & $-1$ & $0$ & $0$\\
$\Sm\pip$ & & & $2$ & $2$ & $0$ & $0$ & $2$ & $0$\\
$\Sz\piz$ & & & & $0$ & $0$ & $-2$ & $1$ & $-1$\\
$\Sz\et$ & & & & & $0$ & $0$ & $\sq{3}$ & $\sq{3}$\\
$\Sp\pim$ & & & & & & $2$ & $0$ & $2$\\
$\Xm\kp$ & & & & & & & $2$ & $-1$\\
$\Xz\kz$ & & & & & & & & $2$\\
\end{tabular}

\subsection{$Q=1$}
\begin{tabular}{c|cccccc}
& $\Dp\kbz$ & $\Dpp\km$ & $\Sz\pip$ & $\Sp\piz$ & $\Sz\et$ & $\Xz\kp$\\
\hline
$\Dp\kbz$ & $1$ & $\sq{3}$ & $-\sq{2}$ & $\sq{\fr{1}{2}}$ & $-\sq{\frac{3}{2}}$
& $0$\\
$\Dpp\km$ & & $3$ & $0$ & $-\sq{\frac{3}{2}}$ & $-\fr{3}{\sq{2}}$ & $0$\\ 
$\Sz\pip$ & & & $0$ & $2$ & $0$ & $\sq{2}$\\
$\Sp\piz$ & & & & $0$ & $0$ & $\sq{2}$\\
$\Sz\et$ & & & & & $0$ & $\sq{6}$\\
$\Xz\kp$ & & & & & & $1$\\
\end{tabular}

\subsection{$Q=2$}
\begin{tabular}{c|cc}
& $\Dpp\kbz$ & $\Sp\pip$\\
\hline
$\Dpp\kbz$ & $0$ & $-\sq{3}$\\
$\Sp\pip$ & & $-2$\\
\end{tabular}

\section{\ul{$S=-2$}}
\subsection{$Q=-2$}
\begin{tabular}{c|cc}
& $\Sm\km$ & $\Xm\pim$\\
\hline
$\Sm\km$ & $-1$ & $-2$\\
$\Xm\pim$ & & $-1$\\
\end{tabular}

\subsection{$Q=-1$}
\begin{tabular}{c|cccccc}
& $\Sm\kbz$ & $\Sz\km$ & $\Xm\piz$ & $\Xm\et$ & $\Xz\pim$ & $\Omm\kz$\\
\hline
$\Sm\kbz$ & $1$ & $\sq{2}$ & $\sq{2}$ & $-\sq{6}$ & $0$ & $0$\\
$\Sz\km$ & & $0$ & $-1$ & $-\sq{3}$ & $-\sq{2}$ & $0$\\
$\Xm\piz$ & & & $0$ & $0$ & $-\sq{2}$ & $-\sq{\frac{3}{2}}$\\
$\Xm\et$ & & & & $0$ & $0$ & $\fr{3}{\sq{2}}$\\
$\Xz\pim$ & & & & & $1$ & $\sq{3}$\\
$\Omm\kz$ & & & & & & $3$\\
\end{tabular}

\subsection{$Q=0$}
\begin{tabular}{c|cccccc}
& $\Sz\kbz$ & $\Sp\km$ & $\Xm\pip$ & $\Xz\piz$ & $\Xz\et$ & $\Omm\kp$\\
\hline
$\Sz\kbz$ & $0$ & $\sq{2}$ & $-\sq{2}$ & $1$ & $-\sq{3}$ & $0$\\
$\Sp\km$ & & $1$ & $0$ & $-\sq{2}$ & $-\sq{6}$ & $0$\\
$\Xm\pip$ & & & $1$ & $\sq{2}$ & $0$ & $\sq{3}$\\
$\Xz\piz$ & & & & $0$ & $0$ & $\sq{\frac{3}{2}}$\\
$\Xz\et$ & & & & & $0$ & $\fr{3}{\sq{2}}$\\
$\Omm\kp$ & & & & & & $3$\\
\end{tabular}

\subsection{$Q=1$}
\begin{tabular}{c|cc}
& $\Sp\kbz$ & $\Xz\pip$\\
\hline
$\Sp\kbz$ & $-1$ & $-2$\\
$\Xz\pip$ & & $-1$\\
\end{tabular}

\section{\ul{$S=-3$}}
\subsection{$Q=-2$}
\begin{tabular}{c|cc}
& $\Xm\km$ & $\Omm\pim$\\
\hline
$\Xm\km$ & $-2$ & $-\sq{3}$\\
$\Omm\pim$ & & $0$\\
\end{tabular}

\subsection{$Q=-1$}
\begin{tabular}{c|cccc}
& $\Xz\km$ & $\Xm\kbz$ & $\Omm\pim$ & $\Omm\et$\\
\hline
$\Xz\km$ & $-1$ & $1$ & $-\sq{\frac{3}{2}}$ & $-\fr{3}{\sq{2}}$\\
$\Xm\kbz$ & & $-1$ & $\sq{\frac{3}{2}}$ & $-\fr{3}{\sq{2}}$\\
$\Omm\pim$ & & & $0$ & $0$\\
$\Omm\et$ & & & & $0$\\
\end{tabular}

\subsection{$Q=0$}
\begin{tabular}{c|cc}
& $\Xz\kbz$ & $\Omm\pip$\\
\hline
$\Xz\kbz$ & $-2$ & $-\sq{3}$\\
$\Omm\pip$ & & $0$\\
\end{tabular}

\section{\ul{$S=-4$}}
\subsection{$Q=-2$}
$\Omm\km\rw\Omm\km~~~~C=-3$

\subsection{$Q=-1$}
$\Omm\kbz\rw\Omm\kbz~~~~C=-3$

\newpage
\section*{Appendix II}
\setcounter{subsection}{0}
\renewcommand{\thesubsection}{\arabic{subsection}}

In the following we write the isospin states which were used to calculate the
amplitudes in the isospin basis. We have used average masses for the isospin
multiplets in the baryon decuplet and pseudoscalar octet; 1232 MeV for the $\De$
$[\Dpp,\Dp,\Dz,\Dm]$, 1385 MeV for $\Sg$ $[\Sp,\Sz,\Sm]$, 1533 MeV for 
$\X$ $[\Xz,\Xm]$, 493 MeV for the $K$ $[\kp,\kz]$ and $\kb$ $[\kbz,\km]$
and 140 MeV for $\pi$ $[\pip,\piz,\pim]$. We took the $\Om$ and $\eta$ masses as
1672 and 547 MeV respectively. In
addition, we use the phase convention $|\pip\rgl=-|1,1\rgl$ and
$|\km\rgl=-|\fr{1}{2},-\fr{1}{2}\rgl$. 

\subsection{\ul{$S=1$}}

$|\De K\ I=1\rgl=\fr{1}{2}\ |\Dz \kz\rgl-\sq{\fr{3}{4}}\ |\Dm\kp\rgl$\\
$|\De K\ I=2\rgl=\sq{\fr{3}{4}}\ |\Dz \kz\rgl+\fr{1}{2}\ |\Dm\kp\rgl$

\subsection{\ul{$S=0$}}

$|\De \pi\ I=\fr{1}{2}\rgl=\sq{\fr{1}{6}}\ |\Dp\pim\rgl-\sq{\fr{1}{3}}\ |\Dz\piz\rgl
-\sq{\fr{1}{2}}\ |\Dm\pip\rgl$\\
$|\Sg K\ I=\fr{1}{2}\rgl=\sq{\fr{1}{3}}\ |\Sz\kz\rgl-\sq{\fr{2}{3}}\ |\Sm\kp\rgl$\\ 
$|\De \pi\ I=\fr{3}{2}\rgl=\sq{\fr{8}{15}}\ |\Dp\pim\rgl-\sq{\fr{1}{15}}\ |\Dz\piz\rgl
+\sq{\fr{2}{5}}\ |\Dm\pip\rgl$\\
$|\Sg K\ I=\fr{3}{2}\rgl=\sq{\fr{2}{3}}\ |\Sz\kz\rgl+\sq{\fr{1}{3}}\ |\Sm\kp\rgl$\\
$|\De \et\ I=\fr{3}{2}\rgl=|\Dz\et\rgl$\\          
$|\De \pi\ I=\fr{5}{2}\rgl=\sq{\fr{3}{10}}\ |\Dp\pim\rgl+\sq{\fr{3}{5}}\ |\Dz\piz\rgl
-\sq{\fr{1}{10}}\ |\Dm\pip\rgl$

\subsection{\ul{$S=-1$}}

$|\Sg \pi\ I=0\rgl=\sq{\fr{1}{3}}\ |\Sp\pim\rgl-\sq{\fr{1}{3}}\ |\Sz\piz\rgl
-\sq{\fr{1}{3}}\ |\Sm\pip\rgl$\\
$|\X K\ I=0\rgl=\sq{\fr{1}{2}}\ |\Xz\kz\rgl-\sq{\fr{1}{2}}\ |\Xm\kp\rgl$\\ 
$|\De \kb\ I=1\rgl=-\sq{\fr{1}{2}}\ |\Dp\km\rgl-\sq{\fr{1}{2}}\ |\Dz\kbz\rgl$\\ 
$|\Sg \pi\ I=1\rgl=\sq{\fr{1}{2}}\ |\Sp\pim\rgl+\sq{\fr{1}{2}}\ |\Sm\pip\rgl$\\
$|\Sg \et\ I=1\rgl=|\Sz\et\rgl$\\
$|\X K\ I=1\rgl=\sq{\fr{1}{2}}\ |\Xz\kz\rgl+\sq{\fr{1}{2}}\ |\Xm\kp\rgl$\\
$|\De \kb\ I=2\rgl=-\sq{\fr{1}{2}}\ |\Dp\km\rgl+\sq{\fr{1}{2}}\ |\Dz\kbz\rgl$\\ 
$|\Sg \pi\ I=2\rgl=\sq{\fr{1}{6}}\ |\Sp\pim\rgl+\sq{\fr{2}{3}}\ |\Sz\piz\rgl
-\sq{\fr{1}{6}}\ |\Sm\pip\rgl$

\subsection{\ul{$S=-2$}}

$|\Sg \kb\ I=\fr{1}{2}\rgl=-\sq{\fr{2}{3}}\ |\Sp\km\rgl-\sq{\fr{1}{3}}\ |\Sz\kbz\rgl$ \\
$|\X \pi\ I=\fr{1}{2}\rgl=\sq{\fr{2}{3}}\ |\Xm\pip\rgl+\sq{\fr{1}{3}}\ |\Xz\piz\rgl$\\
$|\X \et\ I=\fr{1}{2}\rgl=|\Xz\et\rgl$\\
$|\Om K\ I=\fr{1}{2}\rgl=|\Omm\kp\rgl$\\
$|\Sg \kb\ I=\fr{3}{2}\rgl=-\sq{\fr{1}{3}}\ |\Sp\km\rgl+\sq{\fr{2}{3}}\ |\Sz\kbz\rgl$ \\
$|\X \pi\ I=\fr{3}{2}\rgl=-\sq{\fr{1}{3}}\ |\Xm\pip\rgl+\sq{\fr{2}{3}}\ |\Xz\piz\rgl$

\subsection{\ul{$S=-3$}}

$|\X \kb\ I=0\rgl=-\sq{\fr{1}{2}}\ |\Xz\km\rgl-\sq{\fr{1}{2}}\ |\Xm\kbz\rgl$\\
$|\Om \et\ I=0\rgl=|\Omm\et\rgl$\\
$|\X \kb\ I=1\rgl=-\sq{\fr{1}{2}}\ |\Xz\km\rgl+\sq{\fr{1}{2}}\ |\Xm\kbz\rgl$\\
$|\Om \pi\ I=1\rgl=|\Omm\piz\rgl$

\subsection{\ul{$S=-4$}}

$|\Om \kb\ I=\fr{1}{2}\rgl=|\Omm\kbz\rgl$
\end{document}